\documentclass[12pt,prb,superscriptaddress]{revtex4}
\makeatletter
\def\@dotsep{4.5}
\makeatother

\usepackage{amsmath}
\usepackage{amssymb}
\usepackage{graphicx}
\usepackage{ccaption}
\usepackage{psfrag}

\captiondelim{}

\begin{document}

\begin{titlepage}
\title{Reduced-Order Computational Model for the Molecular Dynamics Simulation of Entangled Polymers}
\author{Aruna Mohan}
\affiliation{Department of Chemical Engineering, University of California, Santa Barbara, California 93106, USA}
\author{Glenn H. Fredrickson\footnotemark[1]}
\affiliation{Department of Chemical Engineering, University of California, Santa Barbara, California 93106, USA}
\affiliation{Department of Materials and Materials Research Laboratory, University of California, Santa Barbara, California 93106, USA}
\date{September 8, 2009}

\renewcommand{\thefootnote}{\fnsymbol{footnote}}
\footnotetext[1]{Corresponding author (\tt{ghf@mrl.ucsb.edu})}
\renewcommand{\thefootnote}{\arabic{footnote}}

\begin{abstract}
We present a new reduced-order computational method for the molecular dynamics simulation of entangled polymer systems. The polymer chains are modeled as continuous Gaussian chains, which may interact via interchain and intrachain Lennard--Jones interactions. Our algorithm is based on the application of the molecular dynamics simulation method to the pseudospectral representation of the Fourier modes of the chains. We demonstrate a reduction in computational time from $O(N)$ to $O(1)$ per time step per chain relative to molecular dynamics simulations of entangled polymer systems based on the bead--spring model. We further establish that our model polymer system can exhibit either semiflexible or flexible behavior, depending on whether intrachain excluded volume interactions are present or absent, respectively. Consequently, our model encompasses a wide variety of polymer systems, ranging from semiflexible to flexible polymers.
\end{abstract}

\maketitle
\end{titlepage}

\section{\label{sec:intro} Introduction}

Topological constraints originating from chain connectivity and the noncrossability of chains play a key role in determining the dynamics of polymer chains in many systems of interest, including semidilute and concentrated polymer solutions, entangled polymer melts \cite{degennes,doiedwards} and networks of semiflexible biopolymers. \cite{kas1,mtosh1,mtosh2,mtosh3} Theoretically, the motion of an individual chain in an entangled system has been modeled as being equivalent to the motion of a chain confined within a tube constituted by the surrounding chains. The tube itself is renewed over long time scales as the chain ends diffuse through the network. This mechanism of chain motion is termed reptation.\cite{degennes, doiedwards} The reptation model has been successful in predicting the properties of entangled solutions and melts of flexible polymers,\cite{doiedwards} and has further been extended to theoretically investigate the equilibrium and dynamical properties of semiflexible polymers \cite{odijk, semenov, granek} and membranes.\cite{granek} Moreover, reptation motion has been experimentally visualized in entangled solutions of DNA \cite{perkins} and filamentous actin.\cite{kas1, kas2}

Computer simulations have been extensively employed to verify the predictions of the reptation model and to elucidate the microscopic origins of macroscopically observed properties.\cite{kremergrest, everaers, sukumaran, larson} These studies typically utilize the molecular dynamics simulation method applied to the bead--spring model, with topological constraints enforced via the imposition of repulsive Lennard--Jones interactions among pairs of beads in conjunction with the use of stiff, finitely extensible springs modeled by the FENE (finitely extensible nonlinear elastic) potential to prevent chains from crossing. The molecular dynamics technique applied to the bead--spring model entails the solution of order $N$ equations at each time step for each chain of $N$ beads. Since the relaxation time of a reptating chain of $N$ monomers scales as $N^3$,\cite{degennes, doiedwards} the computational time required to simulate one characteristic relaxation time (tube renewal time) is of order $O(N^4)$ per chain. Further, the use of periodic boundary conditions in simulations requires that the number of chains $M$ scale as $N^{1/2}$ to minimize interactions of a chain with its periodic image upon increasing $N$. Consequently, molecular dynamics simulations based on the bead--spring model prove computationally demanding, thereby constraining the system sizes that can be studied.

While the preceding studies focused mainly on flexible polymer systems, some recent investigations have performed molecular dynamics simulations of the bead--spring model with the inclusion of a bending potential dependent on the angle between adjacent bonds to model the effects of semiflexibility.\cite{faller, auhl, larson2} More recently, a hybrid Brownian dynamics/ Monte Carlo algorithm has been proposed for simulating entangled semiflexible polymer systems.\cite{morse1, morse2} This algorithm is based on the Brownian dynamics of bead--rod chains with a bending potential imposed between adjacent pairs of rods. The bead--rod model approaches the Kratky--Porod wormlike chain in the dual limits of vanishing rod length and an infinite number of rods, while maintaining the contour length fixed. Trial moves generated by the Brownian dynamics algorithm are rejected if they result in the violation of topological constraints and, hence, the time step must be chosen small enough that a majority of the moves is accepted. Computer simulations of entangled semiflexible chains based on the aforementioned methods again prove computationally intensive for large $N$.

In the present contribution, we propose a reduced-order computational model for the molecular dynamics simulation of entangled semiflexible or flexible polymer systems. The chains are modeled as \emph{continuous} Gaussian threads that may interact via interchain and intrachain Lennard--Jones repulsion. The dynamical equations applied to the continuous chain model are solved pseudospectrally in Fourier space using a small number $N_\mathrm{c}$ of collocation points sufficient to accurately resolve the chain dynamics. Remarkably, $N_\mathrm{c}$ is found to be independent of $N$, with $N_\mathrm{c} \ll N$ for long chains of interest. We establish that our model affords a reduction in the order of computational complexity from $O(N)$ to $O(1)$ per time step per chain in relation to molecular dynamics simulations of the bead--spring model. Furthermore, the proposed approach may be readily extended to account for more general interaction potentials, external forces, hydrodynamic interactions in semidilute or concentrated polymer solutions, or to the study of membranes, within a computationally tractable framework.

The paper is organized as follows. We present our model and simulation method in Sec. \ref{sec:model}. Section \ref{sec:saw} contains our results for self-avoiding chains interacting via interchain as well as intrachain excluded volume interactions, while Sec. \ref{sec:phantom} presents results for phantom chains in the absence of intrachain interactions, but with interchain repulsion. The computational advantage of our method relative to molecular dynamics simulations of bead--spring polymer chains is demonstrated in Sec. \ref{sec:compt}. Finally, Sec. \ref{sec:conclude} summarizes our findings and discusses potential extensions and applications of our model.

\section{\label{sec:model} Model}

Our model is based on the application of the molecular dynamics simulation technique to a system of $M$ continuous Gaussian chains, each having chain length $N$, parameterized by the arc length $s \in [0,N]$. The balance of forces acting upon chain $i \ (i=1,..,M)$, represented by the space curve $\mathbf{R}_i(s,t)$, is expressed by the equation
\begin{equation} \label{eq:MDcont}
\frac{\partial^{2}\mathbf{R}_{i}\left( s,t\right) }{\partial t^{2}}=-\Gamma \frac{\partial \mathbf{
R}_{i}\left( s,t\right) }{\partial t}+\mathbf{F}_{i}^\mathrm{elastic}\left( s,t\right) + \mathbf{
F}_{i}^\mathrm{EV}\left( s,t\right) + \mathbf{F}_{i}^\mathrm{thermal}\left( s,t\right) 
\end{equation}
in Lennard--Jones units, where $\Gamma$ denotes the nondimensional segmental friction coefficient. In adopting Lennard--Jones units, we express mass, energy and length in units of the segmental mass $m$ and the Lennard--Jones parameters $\epsilon$ and $\sigma$, respectively. Consequently, the Lennard--Jones time scale $\tau=\sigma(m/\epsilon)^{1/2}$ represents the unit of time, and $\Gamma$ is expressed in units of $m/\tau$. Temperature is measured in units of $\epsilon/k_\mathrm{B}$, where $k_\mathrm{B}$ denotes the Boltzmann constant. The forces appearing on the right hand side of Eq. (\ref{eq:MDcont}) represent the drag force, the elastic restoring force, the net repulsive Lennard--Jones excluded volume force and the thermal force, respectively. The inclusion of drag and thermal forces signifies that the system is coupled to a heat bath, thereby enabling the temperature to be maintained at a preset value and imparting numerical stability to molecular dynamics simulations of Eq. (\ref{eq:MDcont}).

We further introduce the scaled contour variable $\tilde{s} = s/N \in [0,1]$, whereby the net excluded volume force takes the form
\begin{equation} \label{eq:EV}
\mathbf{F}_{i}^\mathrm{EV}(\tilde{s},t) = -N \sum_{j=1}^{M} \int_{0}^{1}d\tilde{s}^{\prime }\frac{\partial }{%
\partial \mathbf{R}_{i}(\tilde{s},t)} v\left( \left| \mathbf{R}_{i}\left( \tilde{s},t \right) -%
\mathbf{R}_{j}\left( \tilde{s}^{\prime },t\right) \right| \right)
\end{equation}
where $v$ denotes the purely repulsive, shifted and truncated Lennard--Jones potential
\begin{equation} \label{eq:LJ}
v(r) = 
\begin{cases}
4  \left[ \left( \frac{ 1 }{r} \right) ^{12}-\left( \frac{1 }{r}\right) ^{6}+\frac{1}{4}\right],  \  r \leq 2^{1/6}  \\
0, \ r > 2^{1/6}
\end{cases}
\end{equation}
The above formulation of excluded volume interactions among continuous chains is particularly well-suited to the study of entangled polymer systems, as the harsh short range repulsions among chains inherently preserve topologically-induced noncrossability constraints. Equation (\ref{eq:EV}), which contains infinite contributions from self-interactions, may be rendered finite by introducing a cutoff $\delta$ such that
\begin{multline} \label{eq:EVreg}
\mathbf{F}_{i}^\mathrm{EV}(\tilde{s},t) = -N 
\sum_{\substack{j=1 \\ j\neq i}}^{M} \int_{0}^{1}d\tilde{s}^{\prime }\frac{\partial }{%
\partial \mathbf{R}_{i}(\tilde{s},t)} v\left( \left| \mathbf{R}_{i}\left( \tilde{s},t\right) -%
\mathbf{R}_{j}\left( \tilde{s}^{\prime },t\right) \right| \right)   \\
 - N \left[  \int_{0}^{\tilde{s}-\delta} d\tilde{s}^{\prime }\frac{\partial }{\partial \mathbf{R}_{i}(\tilde{s},t)} v \left( \left| \mathbf{R}_{i}\left( \tilde{s},t\right) -%
\mathbf{R}_{i}\left( \tilde{s}^{\prime },t\right) \right| \right)  \right.  \\
 + \left. \int_{\tilde{s}+\delta}^{1}d\tilde{s}^{\prime }\frac{\partial }{%
\partial \mathbf{R}_{i}(\tilde{s},t)} v\left( \left| \mathbf{R}_{i}\left( \tilde{s},t\right) -%
\mathbf{R}_{i}\left( \tilde{s}^{\prime },t\right) \right| \right) \right]
\end{multline}
The parameter $\delta$ may be viewed as an adjustable parameter that determines the local chain stiffness and overall chain size, as elaborated in Sec. \ref{sec:saw}.

The elastic restoring force adopts the form \cite{doiedwards}
\begin{equation} \label{eq:elastic}
\mathbf{F}_{i}^\mathrm{elastic}\left( \tilde{s},t\right) = \frac{k}{N^2} \frac{\partial
^{2}\mathbf{R}_{i}(\tilde{s},t)}{\partial \tilde{s}^{2}}
\end{equation}%
derived from the Edwards Hamiltonian of a Gaussian chain, where $k$ represents the nondimensional spring constant and is related to the root mean square segment length $b$ of the Gaussian chain (in the absence of all other interactions) via the definition $k=3k_BT \sigma^2/ (\epsilon b^2)$. The parameter $k$ may, hence, be equivalently interpreted as a dimensionless inverse Lennard--Jones interaction energy. Finally, the thermal force represents white noise with vanishing mean and with the variance \cite{doiedwards} ($i,j=1,..,M$)
\begin{equation} \label{eq:noisevar}
\left\langle \mathbf{F}_{i}^\mathrm{thermal}\left( \tilde{s},t \right) \mathbf{F}_{j}^\mathrm{thermal}\left( \tilde{s}^{\prime },t^{\prime }\right) \right\rangle = \frac{2}{N} T \Gamma \delta _{ij}\delta \left( 
\tilde{s}-\tilde{s}^{\prime }\right) \delta \left( t-t^{\prime }\right) \boldsymbol{\delta }  
\end{equation}
in Lennard--Jones units, where $\boldsymbol{\delta}$ denotes the unit tensor.

An examination of Eqs. (\ref{eq:MDcont})--(\ref{eq:noisevar}) reveals that the chain length $N$ appears only as a parameter in our model. This feature of the continuous chain model stands in contrast to conventional molecular dynamics simulations applied to the bead--spring model, wherein an increase in $N$ necessitates the addition of beads to the chain and the concomitant solution of additional dynamical equations, and may be exploited in devising a reduced-order simulation method as detailed below.

We begin by decomposing the chain coordinates into Fourier cosine (i.e., ``Rouse'') modes by means of the transformation
\begin{equation} \label{eq:cosmodes}
\mathbf{R}_{i}\left( \tilde{s},t\right)
= \sum_{p=0}^{N_\mathrm{c}-1} \hat{\mathbf{R}}_{i}^{p}( t) \cos \left( \tilde{s} p\pi \right)
\end{equation}
where we have truncated the Fourier series after $N_\mathrm{c}$ terms, and $\hat{\mathbf{R}}_{i}^{p}(t), \ p=0,..,N_\mathrm{c}-1 $ denote the $N_\mathrm{c}$ lowest cosine modes of the chain. The choice of cosine modes is motivated by the free chain boundary conditions, $\partial \mathbf{R}_i(\tilde{s},t)/\partial \tilde{s} = \mathbf{0}$ at $\tilde{s}=0,\ 1$. Finally, substitution of Eq. (\ref{eq:cosmodes}) into Eq. (\ref{eq:MDcont}) in conjunction with the orthogonality of the cosine modes yields the equation
\begin{equation} \label{eq:MDcos}
\frac{\partial^2 \hat{\mathbf{R}}_i^q (t) } {\partial t^2}=-\Gamma \frac{\partial \hat{
\mathbf{R}}_i^q(t)} {\partial t} - \frac{k}{N^2} q^2 \pi^2 \hat{\mathbf{R}}_i^q(t) + 
\hat{\mathbf{F}}_i^{\mathrm{EV},\; q}(t) + \hat{\mathbf{F}}_i^{\mathrm{thermal},\; q}(t)
\end{equation}
for $q=0,..,N_\mathrm{c}-1$, where $\hat{\mathbf{F}}_i^{\mathrm{EV},\; q}(t)$ are cosine transforms of $\mathbf{F}_i^\mathrm{EV}(\tilde{s},t)$ defined in analogy with Eq. (\ref{eq:cosmodes}) in the following manner:
\begin{equation} \label{eq:forcecosmodes}
\mathbf{F}_{i}^\mathrm{EV} \left( \tilde{s},t\right)
= \sum_{p=0}^{N_\mathrm{c}-1} \hat{\mathbf{F}}_{i}^{\mathrm{EV}, \; p}(t) \cos \left( \tilde{s} p\pi \right)
\end{equation}
The transformed thermal force terms $\hat{\mathbf{F}}_i^{\mathrm{thermal},\; q}(t)$ have vanishing mean and the variance 
\begin{equation} \label{eq:fouriernoisvar}
\left\langle \hat{\mathbf{F}}_i^{\mathrm{thermal},\; p}(t) \hat{\mathbf{F}}_j^{\mathrm{thermal},\; q} (t^\prime) \right\rangle = \frac{2}{N} (2-\delta_{q0}) T \Gamma \delta _{ij}\delta(t-t^\prime) \delta _{pq} \boldsymbol{\delta}
\end{equation}
in Lennard--Jones units.

We implement the transformations between the real space variables $\mathbf{R}_{i}\left( \tilde{s},t\right)$ and $\mathbf{F}_{i}^\mathrm{EV} \left( \tilde{s},t\right)$ and their corresponding Fourier space representations pseudospectrally at the $N_\mathrm{c}$ collocation points $\tilde{s}=l/(N_\mathrm{c}-1), \ l = 0,.., N_\mathrm{c}-1$. Each such discrete Fourier cosine transformation is performed in $O(N_\mathrm{c} \log N_\mathrm{c})$ time by means of the Fast Fourier Transform algorithm.\cite{frigo} Thus, at each time step, for a system of $M$ chains, our algorithm involves computation of the excluded volume forces in real space in $O(M N_\mathrm{c})$ time on average using neighbor lists, followed by a transformation of the forces to Fourier space in $O(M N_\mathrm{c} \log N_\mathrm{c})$ time, the subsequent time-stepping of the dynamical equations for the $N_\mathrm{c}$ Fourier modes on each chain in $O(M N_\mathrm{c})$ time and, finally, the transformation of the updated Fourier coordinates to real space in $O(M N_\mathrm{c} \log N_\mathrm{c})$ time. The overall time complexity of our algorithm is $O(N_\mathrm{c} \log N_\mathrm{c})$ per chain per time step. The computational advantage of this method stems from the fact that there is no explicit dependence of $N_\mathrm{c}$ on $N$. We demonstrate in Secs. \ref{sec:saw} and \ref{sec:phantom} that convergence is achieved with $N_\mathrm{c}$ independent of $N$ and $N_\mathrm{c} << N$ for chain lengths $N$ of interest. Therefore, our algorithm engenders a reduction in computational time from $O(N)$ to $O(1)$ per time step per chain in comparison with molecular dynamics simulations of bead--spring chains.

We performed simulations based on the above method for systems of sizes $N=80$ with $M=20$ chains, $N=100$ with $M=20$ chains,  $N=150$ with $M=20$ chains, $N=200$ with $M=32$ chains, and $N=1\:000$ with $M=72$ chains in a cubic simulation box with periodic boundaries at a monomer number density of $0.85$, weakly coupled to a heat bath at a temperature of $T=1$ with a friction constant of $\Gamma=0.5$.\cite{kremergrest} These system sizes are comparable to those studied by Kremer and Grest,\cite{kremergrest} with $M$ chosen to scale with $N^{1/2}$. \cite{kremergrest} The velocity Verlet algorithm was adopted for the integration of the dynamical equations [Eq. (\ref{eq:MDcos})], with a time step of $\Delta t=0.006\tau$.\cite{kremergrest} Equilibrated chain configurations were generated based on the ``fast push-off'' procedure of Ref. \onlinecite{auhl}, starting from initial freely jointed chain conformations with the step length connecting adjacent collocation points determined by assuming Gaussian statistics for the chain size, followed by sufficiently long runs with the Lennard--Jones potential to ensure equilibration. The integrals involved in the computation of the excluded volume force in Eq. (\ref{eq:EVreg}) were performed using the composite trapezoidal rule, which was found to yield convergence with fewer collocation points than the composite Simpson's rule.\cite{press} Our results are presented in Sec. \ref{sec:saw} for self-avoiding chains, and in Sec. \ref{sec:phantom} for phantom chains.

\section{\label{sec:saw} Self-Avoiding Chains}

In this section, we present results for self-avoiding chains in the presence of both interchain and intrachain excluded volume interactions. As intimated in Sec. \ref{sec:model}, self-interactions are prohibited by introducing a short distance cutoff $\Delta \tilde{s} = \delta$ (or, equivalently, $\Delta s=N\delta$), defined in Eq. (\ref{eq:EVreg}) as the fraction of chain segments on each side of a given point on the chain that does not interact with it via Lennard--Jones forces. The number of collocation points $N_\mathrm{c}$ required for convergence of the integrals in Eq. (\ref{eq:EVreg}) is, hence, a function of $\delta$. We select $\delta$ so as to reproduce the mean square end-to-end distance obtained from comparable bead--spring simulations of flexible chains. \cite{auhl} We will initially consider chains with $k=75$. For a chain length of $N=100$, the value $\delta=0.05$ is found to yield a mean square end-to-end distance of $\left\langle R^2\right\rangle \approx 1.5N$, in accord with values cited in Ref. \onlinecite{auhl} for flexible chains. Larger values of $\delta$ were found to cause a shrinking of the chain, while smaller values resulted in an increase in $\left\langle R^2\right\rangle$.

\begin{figure}
\centering
\resizebox{80mm}{!}{\includegraphics{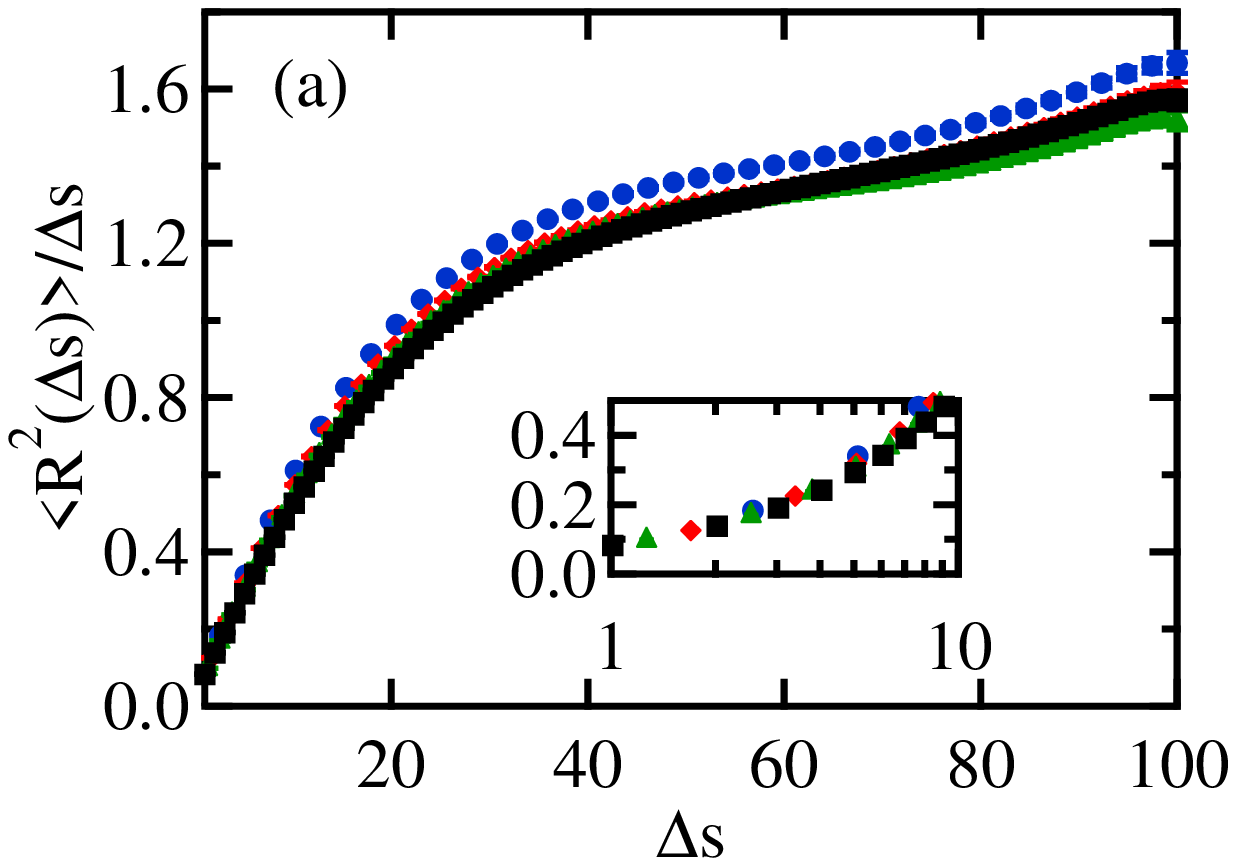}}
\vskip 1.25em
\centering
\resizebox{80mm}{!}{\includegraphics{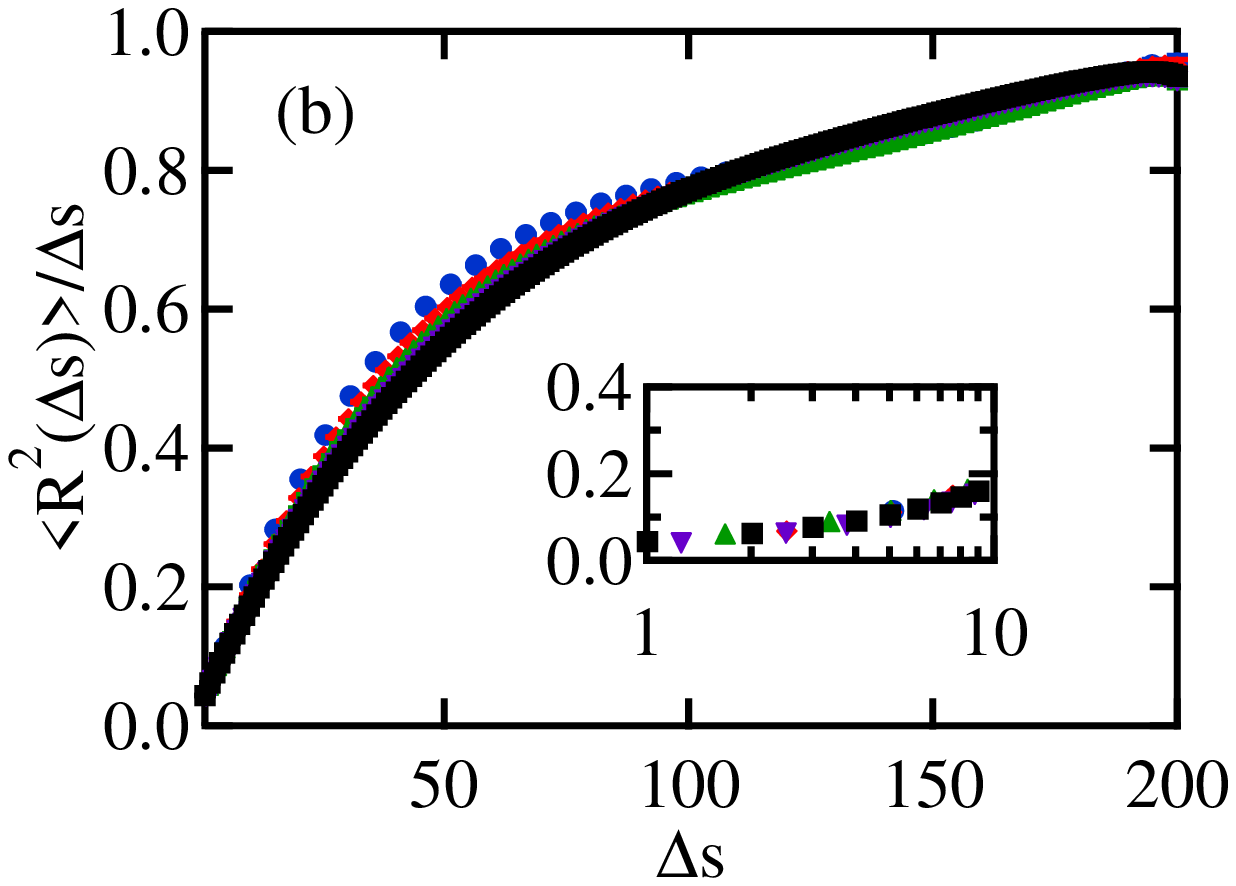}}
\caption[(Color online) Mean square internal distance $\left\langle R^2(\Delta s) \right\rangle$ between points separated by arc length $\Delta s$, normalized by $\Delta s$, for self-avoiding chains of length (a) $N=100$ with $N_\mathrm{c}=40$ (circles), $60$ (diamonds), $80$ (triangles) and $100$ (squares), and (b) $N=200$ with $N_\mathrm{c}=40$ (circles), $80$ (diamonds), $120$ (triangles), $160$ (inverted triangles) and $200$ (squares), and with $k=75$. The error bars represent the standard error. Inset: The region $\Delta s < N\delta$ exhibits near Gaussian scaling.] {\label{fig:msid100200}}
\end{figure}

Figure \ref{fig:msid100200} illustrates the mean square internal distance $\left\langle R^2(\Delta s)\right\rangle$ between points separated by an arc length of $\Delta s$ along the chain contour for self-avoiding chains of lengths $N=100$ and $N=200$, normalized by $\Delta s$. Averages are performed over all segments of size $\Delta s$ for all $M$ chains in each system, and over approximately $100$ equilibrated system configurations separated by time intervals of at least $10^4\tau$ to ensure statistical independence. It is evident that convergence is achieved with $N_\mathrm{c}=40$. Results for chain lengths $N=80,\ N=150$ and $N=1\:000$ (not presented here) also converge with $N_\mathrm{c}=40$ collocation points. The inset of Fig. \ref{fig:msid100200} reveals that the chain statistics remain approximately Gaussian for $\Delta s < N\delta$, where self-interactions are prohibited. Over intermediate scales such that $N\delta < \Delta s < N$, the chain statistics approach rodlike behavior, with $\left\langle R^2 (\Delta s)\right\rangle \sim (\Delta s)^2$. This observation suggests the presence of correlations along the backbone of a continuous chain, actuated by repulsive intrachain interactions. The decrease in $\left\langle R^2(N)\right\rangle/N$ with increase in $N$ may reflect the increase in the number of segments $N\delta$ excluded from self-interactions as $N$ increases.

\begin{figure}
\centering
\psfrag{ucorr}[][]{\LARGE $\left\langle \mathbf{u}(\Delta \text{s})\cdot\mathbf{u}(0) \right\rangle$}
\resizebox{80mm}{!}{\includegraphics{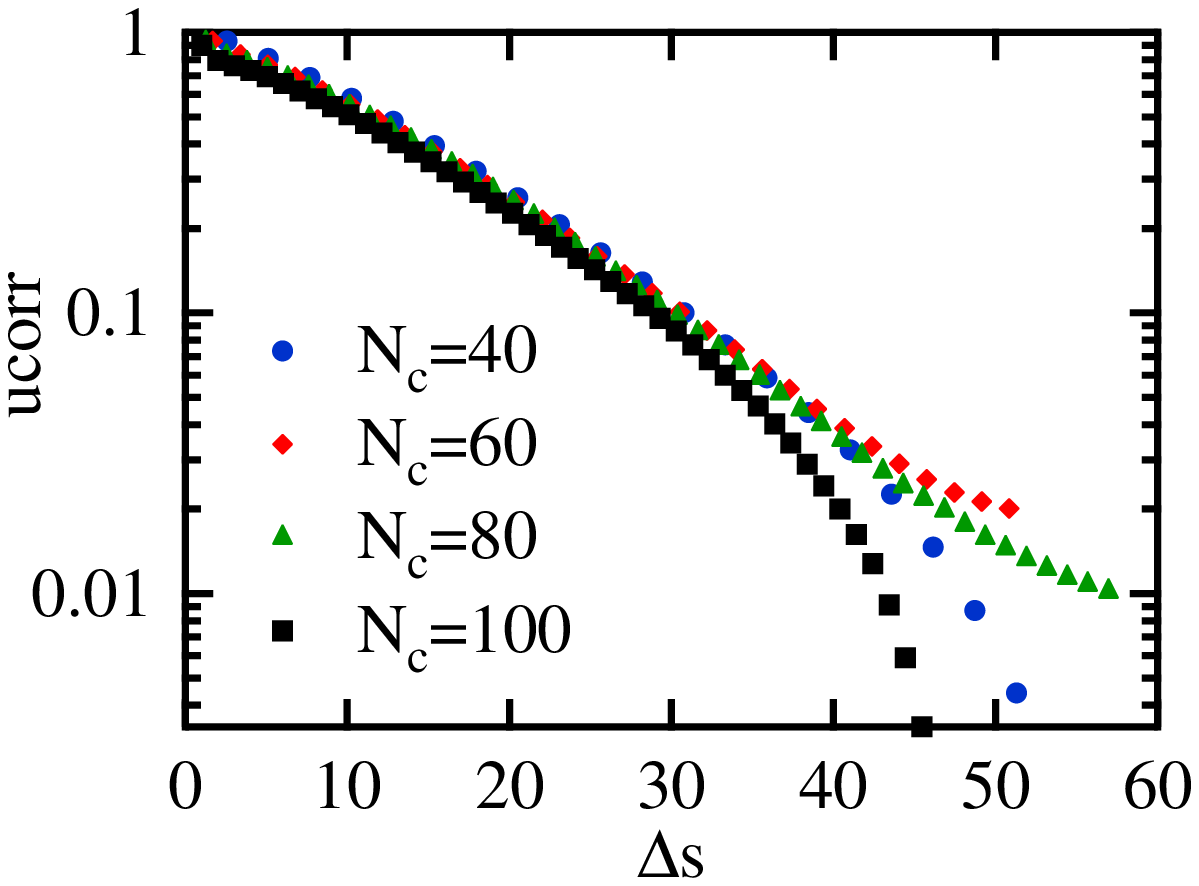}}
\caption[(Color online) Correlation function of the tangent vector $\mathbf{u}$ at points separated by $\Delta s$, $<\mathbf{u}(\Delta s)\cdot\mathbf{u}(0)>$, for self-avoiding chains of length $N=100$ using $N_\mathrm{c}=40,\ 60,\ 80$ and $100$ collocation points, with $k=75$.]{\label{fig:uc100}}
\end{figure}

The correlation function of the unit tangent $\mathbf{u}(s,t)=\partial \mathbf{R}(s,t)/\partial s/\left|\partial \mathbf{R}(s,t)/\partial s\right|$ to the chain contour $\mathbf{R}(s,t)$, $\left\langle \mathbf{u}(\Delta s)\cdot \mathbf{u}(0) \right\rangle$, is plotted as a function of $\Delta s$ in Fig. \ref{fig:uc100} for a chain of length $N=100$. The averaging is performed over points at a separation of $\Delta s$ for equilibrated chains at time intervals of $60\tau$ over a time period of $12\:000\tau$. The behavior seen in Fig. \ref{fig:uc100} over the region $N\delta < \Delta s < N$ is consistent with the relation $\left\langle \mathbf{u}(s)\cdot \mathbf{u}(0) \right\rangle \sim \exp(-s/\lambda)$ for a wormlike chain.\cite{doiedwards} The parameter $\lambda$ denotes the persistence length measured in terms of number of segments, and is found to be approximately $15\%$ of the chain contour for $N=100$. For $\Delta s < N\delta$, the chain conformation resembles an ideal random walk with uncorrelated steps. Over large scales such that $\Delta s \gg \lambda$, the chain statistics approach Gaussian behavior, indicated by a flattening of the $\left\langle R^2 (\Delta s) \right\rangle/\Delta s$ curves in Fig. \ref{fig:msid100200} as $\Delta s$ approaches $N$.

\begin{figure}
\centering
\resizebox{80mm}{!}{\includegraphics{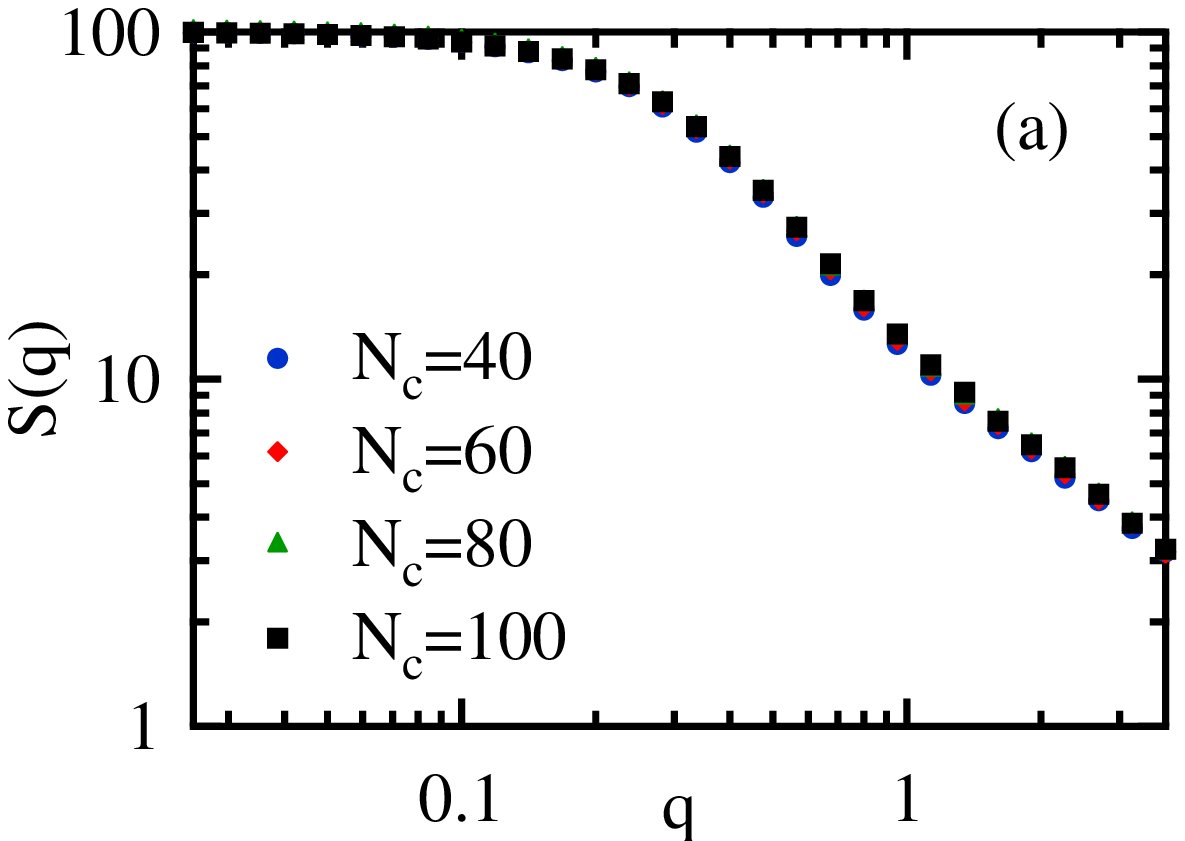}}
\vskip 1.25em
\centering
\resizebox{80mm}{!}{\includegraphics{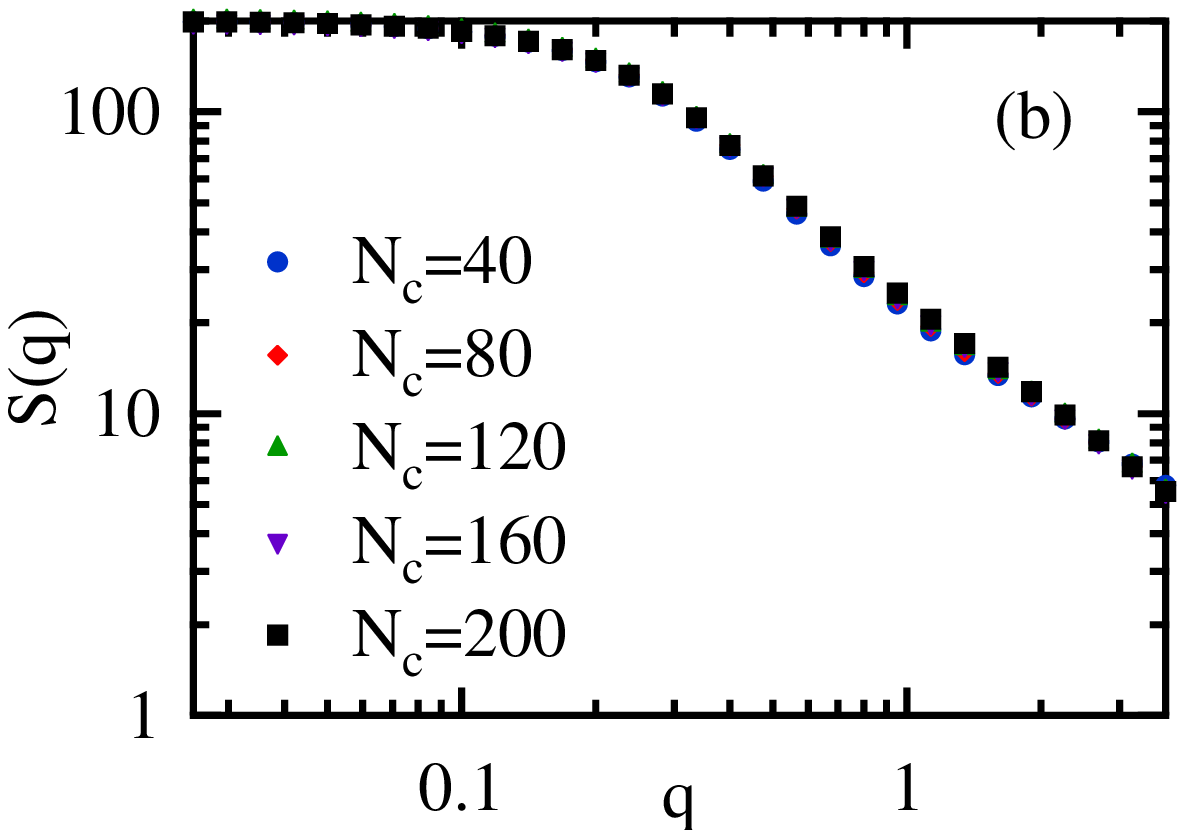}}
\caption[(Color online) Structure factor $S(q)$ for self-avoiding chains of length (a) $N=100$ using $N_\mathrm{c}=40,\ 60,\ 80$ and $100$ collocation points, and (b) $N=200$ using $N_\mathrm{c}=40,\ 80,\ 120,\ 160$ and $200$ collocation points, with $k=75$.]{\label{fig:sq100200}}
\end{figure}

The results presented in the remainder of this section are based on simulation lengths of $108\:000\tau$ for $N=80,\ 100,\ 150$ and $200$, and $216\:000\tau$ for $N=1\:000$. Numerical integration, wherever required, is performed using the composite trapezoidal rule.\cite{press} Figure \ref{fig:sq100200} depicts the coherent structure factor \cite{tildesley} $S(q) = \left\langle \rho_\mathbf{q} \rho_{-\mathbf{q}} \right\rangle/N $ for $N=100$ and $200$, where $\rho_\mathbf{q}$ denotes the Fourier transform of the density 
\begin{equation} \label{eq:ccrho}
\rho(\mathbf{r},t) = \int_0^N ds \; \delta^3(\mathbf{r} - \mathbf{R}(s,t))
\end{equation}
of a continuous chain at a given $t$ with wave vector $\mathbf{q}$. Averaging is performed over $20$ randomly chosen $\mathbf{q}$ vectors at each magnitude $q$ for all chains in equilibrated configurations equally spaced in time at intervals of $1\:080\tau$. As before, convergence is achieved with $N_\mathrm{c}=40$. The slope of $\log S(q)$ vs $\log q$ in the linear region corresponding to $q > 2\pi/\left\langle R^2 \right\rangle^{1/2}$ is found to be $-1.3 \ \pm \ 0.05$ for $N=80,\ 100,\ 150$ and $200$. A slope of $-1.5 \ \pm \ 0.05$ is found in the case of $N=1\:000$. These results indicate behavior intermediate between rigid rods $[S(q) \sim 1/q]$ and flexible chains $[S(q) \sim 1/q^2]$.\cite{doiedwards} The chain statistics are expected to revert to Gaussian behavior over length scales below the short distance cutoff.

\begin{figure}
\centering
\resizebox{80mm}{!}{\includegraphics{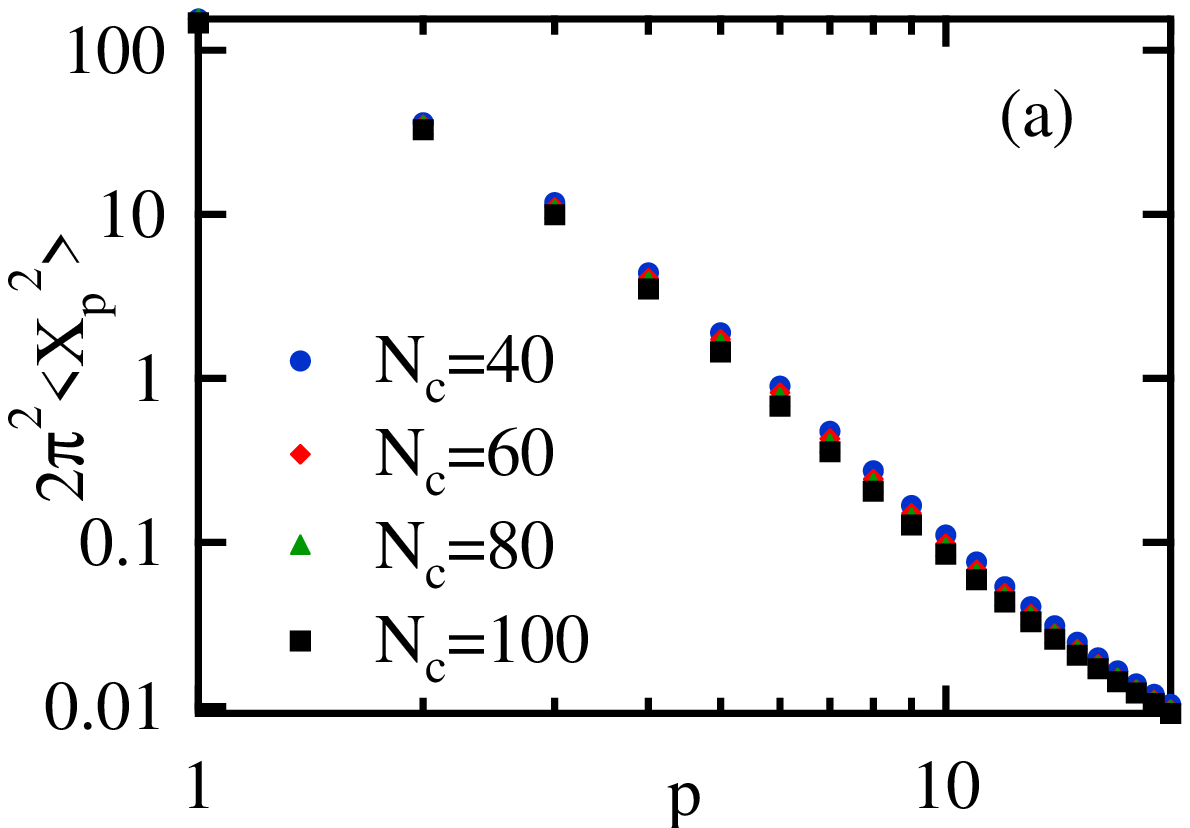}}
\vskip 1.25em
\centering
\resizebox{80mm}{!}{\includegraphics{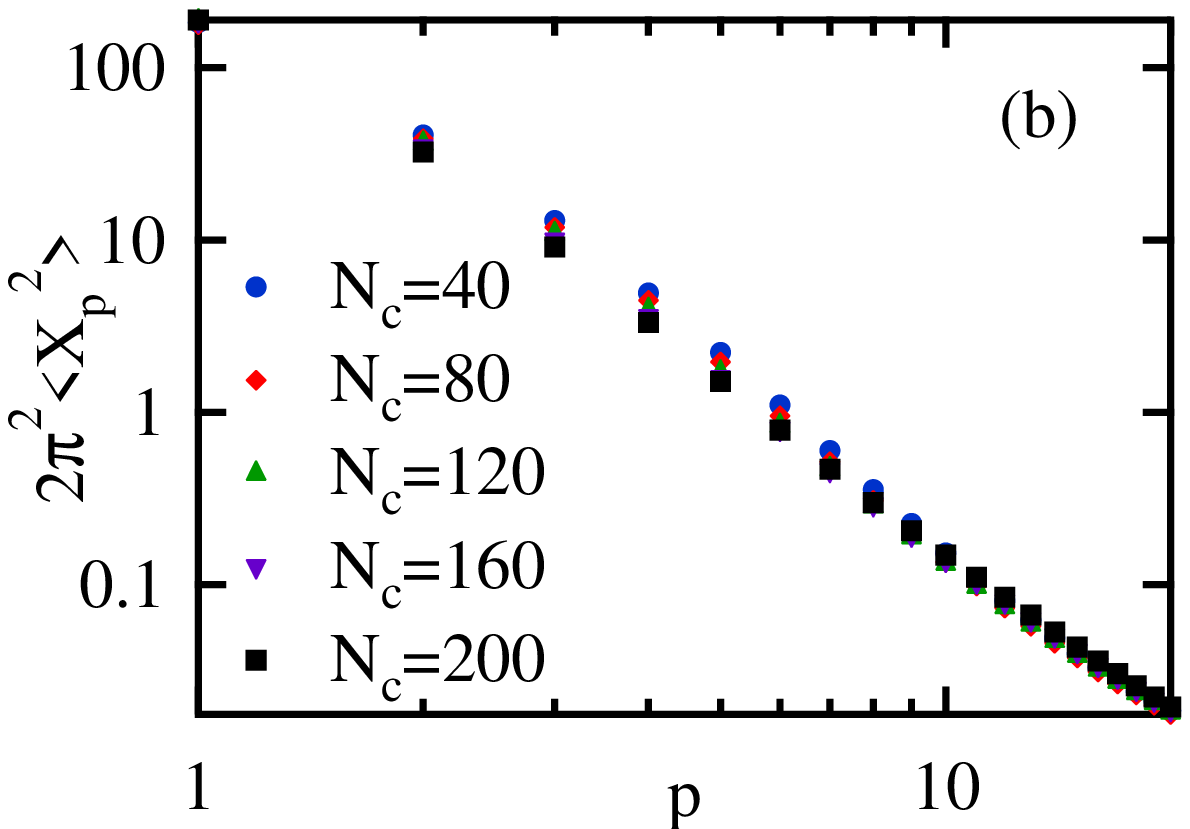}}
\caption[(Color online) Amplitude $<X_p(0)^2>$ of the $p^\mathrm{th}$ Rouse mode, normalized such that the data for $p=1$ coincide with the chain end-to-end distance for a Rouse chain, plotted as a function of $p$ for self-avoiding chains of length (a) $N=100$ using $N_\mathrm{c}=40,\ 60,\ 80$ and $100$ collocation points, and (b) $N=200$ using $N_\mathrm{c}=40,\ 80,\ 120,\ 160$ and $200$ collocation points, with $k=75$.]{\label{fig:xpsq100200}}
\end{figure}

We next consider the behavior of the amplitudes $\left\langle X_p (0)^2 \right \rangle$ of the $p^\mathrm{th}$ Fourier cosine mode, defined by the expression \cite{doiedwards}
\begin{equation} \label{eq:rousemodes}
\mathbf{X}_p(t) = \frac{1}{N} \int_0^N ds \cos\left( \frac{p\pi s}{N} \right) \mathbf{R}(s,t)
\end{equation}
with $2\pi^2\left\langle X_p (0)^2 \right \rangle = \left\langle R^2 \right \rangle /p^2$ for a noninteracting (Rouse) chain at equilibrium. Figure \ref{fig:xpsq100200} depicts the normalized mode amplitudes as a function of $p$ for $N=100$ and $200$, averaged over all chains with configurations sampled at intervals of $60\tau$. The results exhibit the scaling $\left\langle X_p (0)^2 \right \rangle \sim 1/p^\alpha$ for $1 < p < 20$, with $\alpha=3.9 \ \pm \ 0.07$ for $N=100$ and $\alpha=3.7 \ \pm \ 0.07$ for $N=200$.  These observations are in accord with the behavior obtained from the wormlike chain bending energy \cite{doiedwards}
\begin{equation} \label{eq:wlcH}
\mathcal{H} = \frac{1}{2} k_\mathrm{B}T\lambda \int ds \left| \frac{\partial^2 \mathbf{R}}{\partial s^2} \right|^2
\end{equation}
which yields $\left\langle X_p (0)^2 \right \rangle \sim 1/p^4$ from the equipartition of energy among the Fourier modes at equilibrium. The values $\alpha=4.0 \ \pm \ 0.07$, $\alpha=3.8 \ \pm \ 0.07$, and $\alpha=2.6 \ \pm \ 0.03$ are observed for $N=80,\ 150$ and $1\:000$, respectively, indicating closer agreement with Rouse behavior as $N$ is increased.

\begin{figure}
\centering
\resizebox{80mm}{!}{\includegraphics{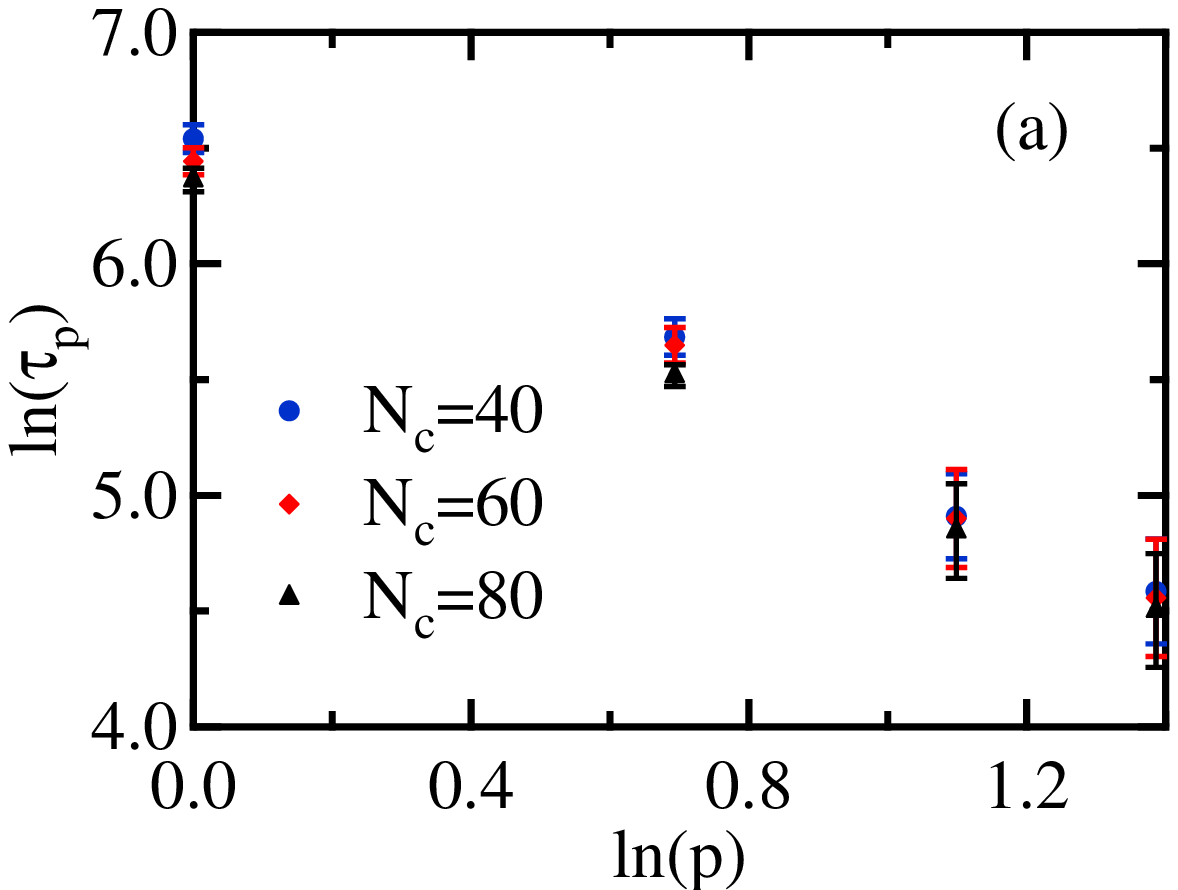}}
\vskip 1.25em
\centering
\resizebox{80mm}{!}{\includegraphics{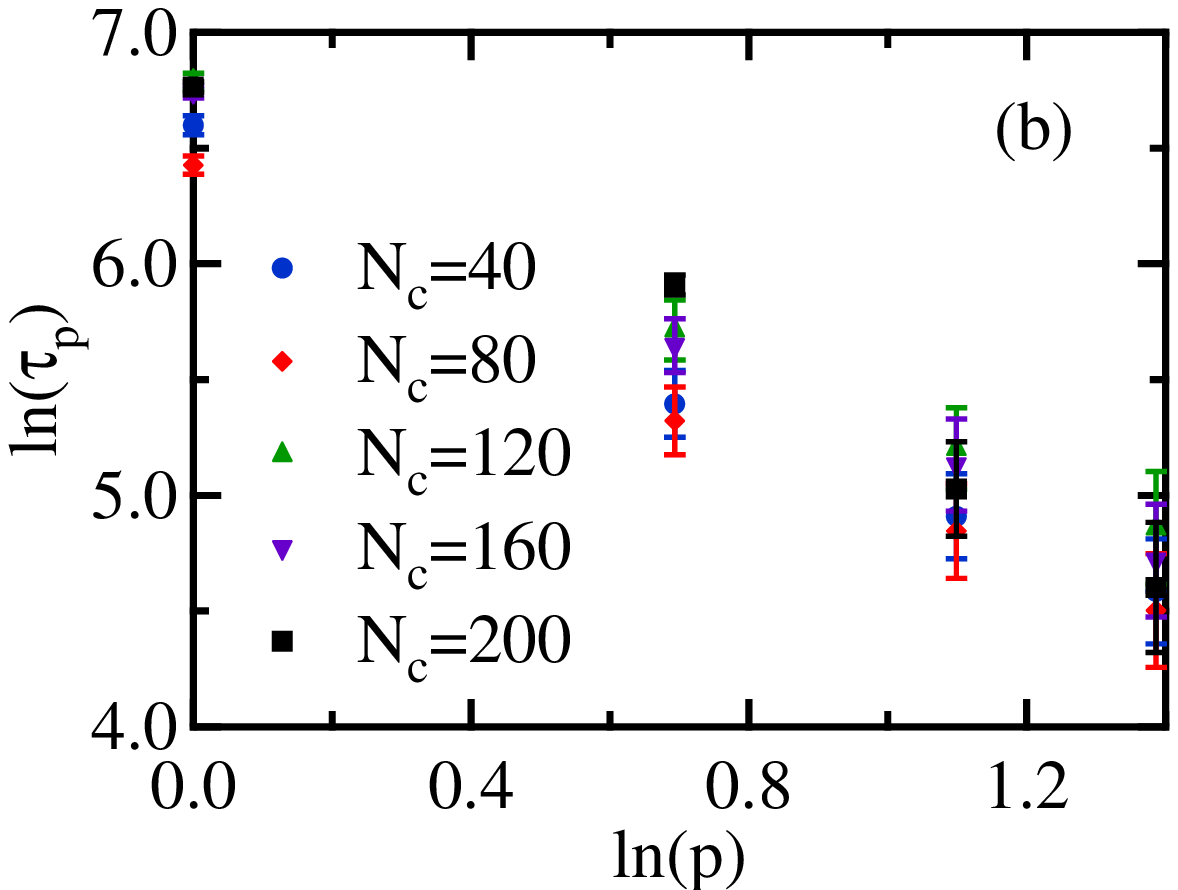}}
\caption[(Color online) Mode correlation time $\tau_p$ of the $p^\mathrm{th}$ Rouse mode plotted as a function of $p$ for self-avoiding chains of length (a) $N=80$ using $N_\mathrm{c}=40,\ 60$ and $80$ collocation points, and (b) $N=200$ using $N_\mathrm{c}=40,\ 80,\ 120,\ 160$ and $200$ collocation points, with $k=75$. The error bars are derived from $95\%$ confidence bounds on the linear regression of $\ln \left\langle \mathbf{X}_p(t) \cdot \mathbf{X}_p(0) \right\rangle$ on $t$.]{\label{fig:tau80200}}
\end{figure}

The Fourier modes defined by Eq. (\ref{eq:rousemodes}) coincide with the normal coordinates of a Rouse chain.\cite{doiedwards} However, the normal modes of a wormlike chain are not identical to the Fourier modes.\cite{pecora,brang} Nonetheless, since mode coupling effects are weak, the relaxation times of the slow Fourier modes, $\tau_p$, may still be determined assuming single exponential relaxation by employing the relation $\left\langle \mathbf{X}_p(t) \cdot \mathbf{X}_p(0) \right\rangle \sim \exp(-t/\tau_p)$.\cite{brang} A single wormlike chain at equilibrium possesses the relaxation spectrum $\tau_p \sim 1/p^4$. However, the relaxation dynamics of semiflexible polymers are damped by internal friction arising in entangled polymer systems, as demonstrated in Ref. \onlinecite{poirier}. These authors predict a transition from $\tau_p \sim 1/p^4$ to $\tau_p \sim \mathrm{const}$ with increasing $p$. Figure \ref{fig:tau80200} depicts the mode relaxation times of the four slowest modes for $N=80$ and $N=200$. The relaxation of faster modes with $p > 4$ does not remain single exponential over the times considered, owing to mode mixing with the slower modes, and the corresponding relaxation times could not be determined accurately. A linear fit of $\ln \tau_p$ vs $\ln p$ for $p \leq 4$ yields an exponent of $1.4 \ \pm \ 0.2$ for $N=80$ and $1.6 \ \pm \ 0.2$ for $N=200$. These results may indicate a transition between the low-$p$ and high-$p$ regimes predicted in Ref. \onlinecite{poirier}.

\begin{figure}
\centering
\resizebox{80mm}{!}{\includegraphics{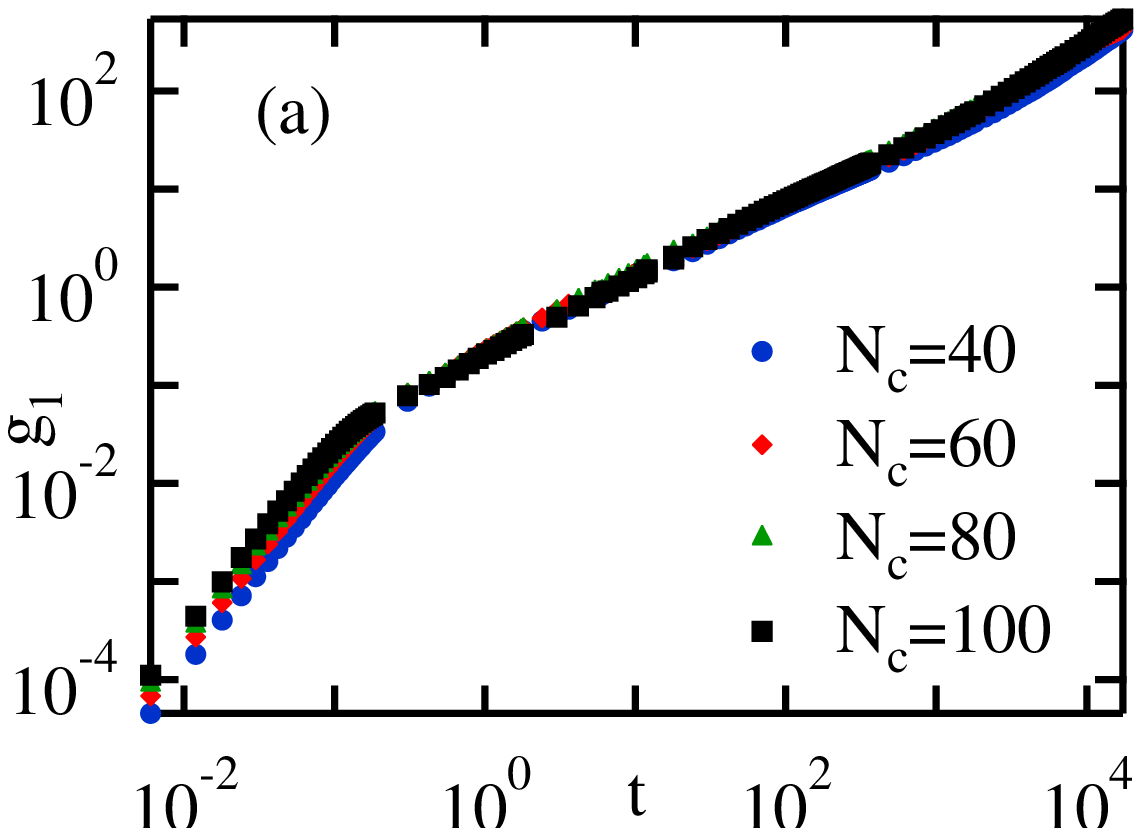}}
\vskip 1.25em
\centering
\resizebox{80mm}{!}{\includegraphics{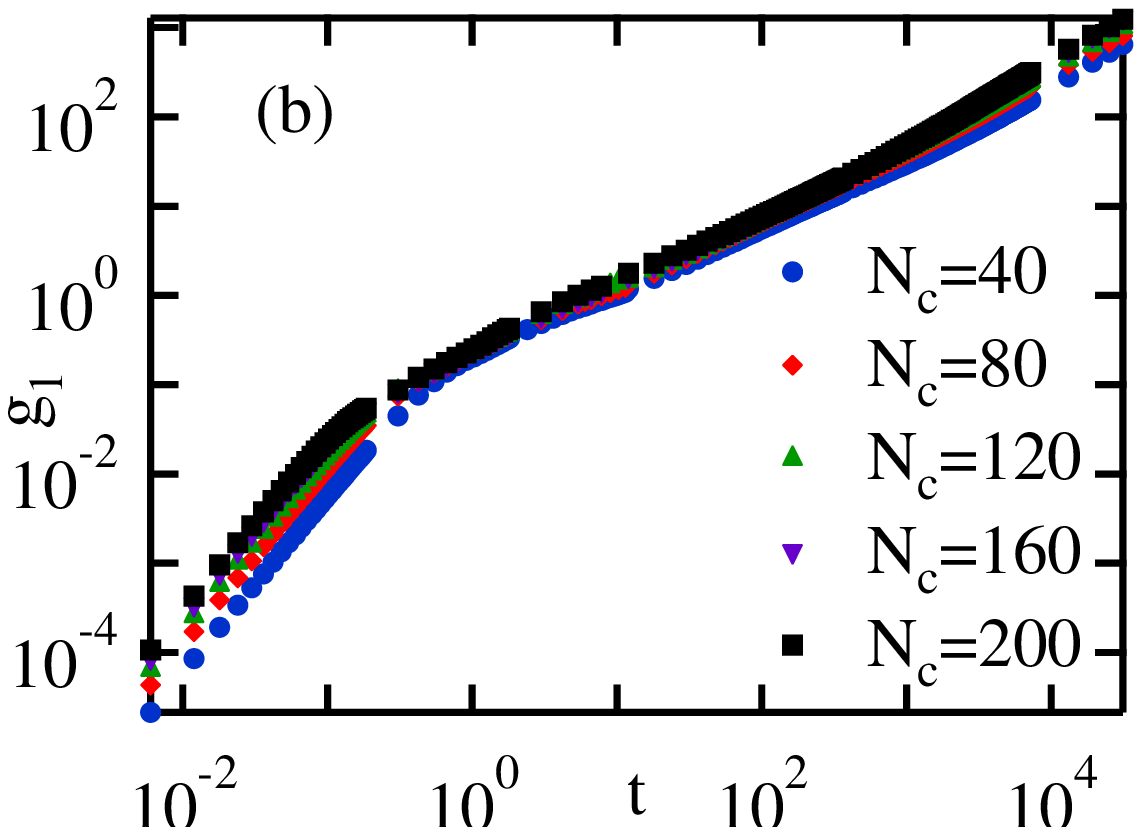}}
\caption[(Color online) Mean square displacement $g_1(t)$ at time $t$ averaged over the innermost $5\%$ of the chain for self-avoiding chains of length (a) $N=100$ using $N_\mathrm{c}=40,\ 60,\ 80$ and $100$ collocation points, and (b) $N=200$ using $N_\mathrm{c}=40,\ 80,\ 120,\ 160$ and $200$ collocation points, with $k=75$.]{\label{fig:msd100200}}
\end{figure}

Finally, we consider the behavior of the mean square displacement $g_1(t) = \left\langle \left| \mathbf{R}(s,t)-\mathbf{R}(s,0) \right|^2 \right\rangle$ averaged over the innermost $5\%$ of the chain, which is strongly influenced by topological constraints.\cite{kremergrest} The coordinates are calculated with respect to the center of mass of the system to eliminate the effect of overall system diffusion arising from the Langevin terms in the equations of motion.\cite{kremergrest} Averaging is performed over at most a third of the total simulation time. Figure \ref{fig:msd100200} illustrates our results for $N=100$ and $N=200$. At very early times, a ballistic regime exhibiting the scaling behavior $g_1(t) \sim t^2$ is observed, which may be attributed to correlated motions of chain segments during inertial relaxation. Over long time scales, the behavior approaches normal diffusion, while at intermediate time scales, subdiffusive behavior is observed. The scaling exponents in this region are $0.80 \ \pm \ 0.01$, $0.75 \ \pm \ 0.01$, $0.74 \ \pm \ 0.01$, $0.71 \ \pm \ 0.01$ and $0.61 \ \pm \ 0.01$ for $N=80,\ 100,\ 150,\ 200$ and $1\:000$, respectively. Our results for short chains are similar to those from prior simulations of a needle in a planar obstacle course, \cite{frey} where a transition from ballistic to diffusive behavior was observed at low densities of obstacles. A plateau in the mean square displacement preceding diffusive behavior was observed in Ref. \onlinecite{frey} at high obstacle densities, indicating tube confinement and eventual escape of the needle. The behavior $g_1 \sim t^{0.75}$ has been predicted for entangled semiflexible chains having persistence length intermediate between the entanglement length and the contour length over time scales shorter than the entanglement time.\cite{granek} However, the exclusion of self-interactions results in Gaussian statistics for $\Delta s < N \delta$ in our system, and consequently, the chains exhibit semiflexible behavior only over intermediate scales. The increase in the short distance cutoff $N \delta$ with increase in $N$ may be responsible for the closer agreement with Rouse behavior as $N$ is increased.

\begin{figure}
\centering
\resizebox{80mm}{!}{\includegraphics{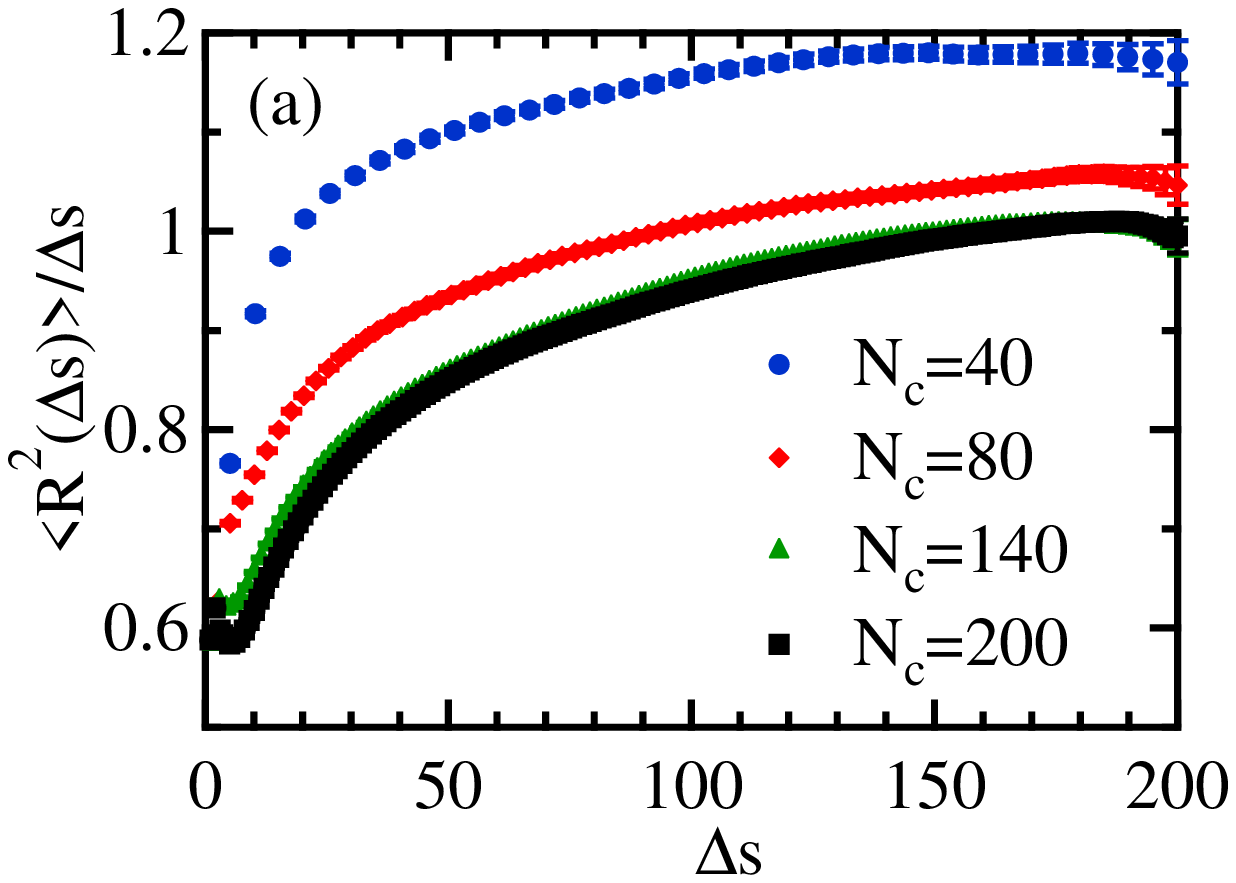}}
\vskip 1.25em
\centering
\resizebox{80mm}{!}{\includegraphics{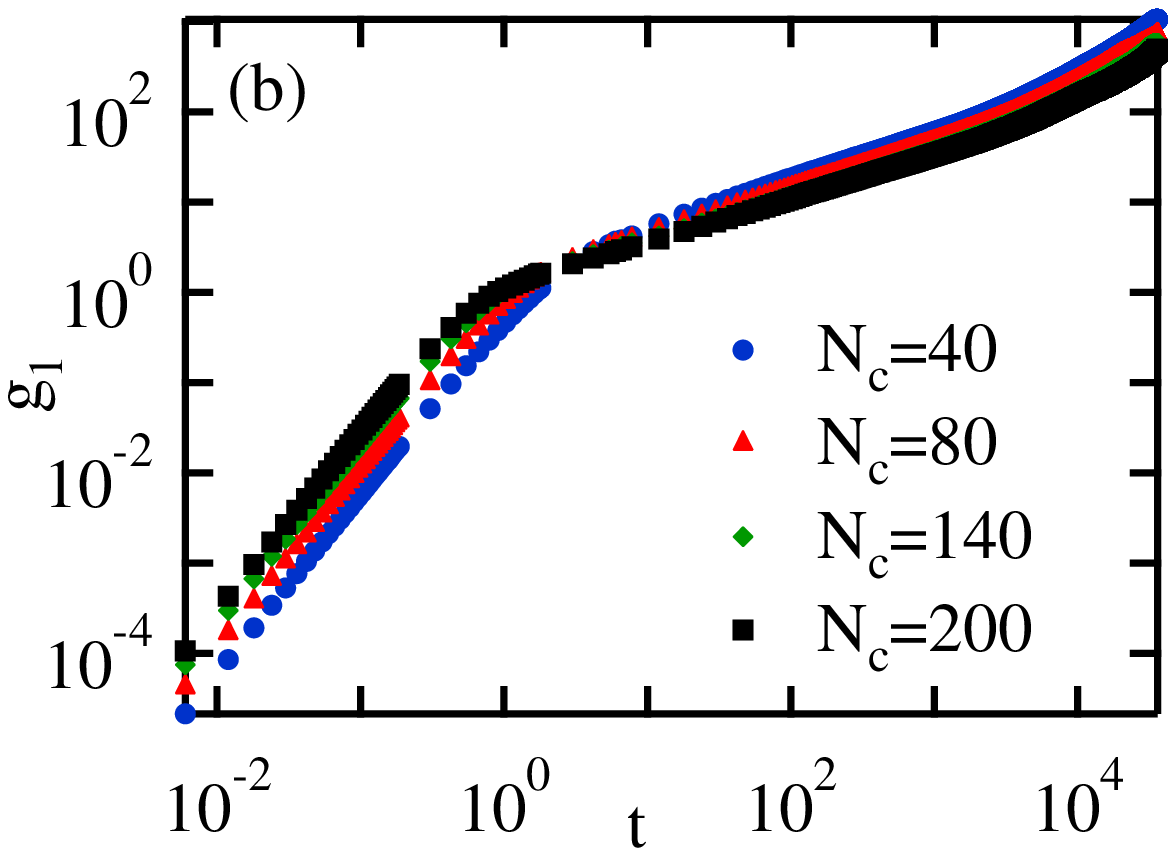}}
\caption[(Color online) Normalized mean square internal distance $\left\langle R^2(\Delta s) \right\rangle/\Delta s$ (a) and mean square displacement $g_1(t)$ averaged over the innermost $5\%$ of the chain (b) for self-avoiding chains of length $N=200$ using $N_\mathrm{c}=40,\ 80,\ 140$ and $200$ collocation points, with $k=3$. The error bars in (a) represent the standard error.]{\label{fig:k3200}}
\end{figure}

The effect of decreasing $k$ is illustrated in Fig. \ref{fig:k3200}, which depicts the normalized mean square internal distance $\left\langle R^2(\Delta s)\right\rangle/\Delta s$ averaged over samples taken at time intervals of $2\:160\tau$, which exceeds the longest relaxation time, for each chain in the system, and the mean square displacement $g_1(t)$, both for $N=200$ with $k=3$. In accord with Rouse behavior in the unentangled regime, subsequent to the ballistic regime at very early times, $g_1(t)$ exhibits a transition from the Rouse scaling of $t^{0.5}$ at intermediate times to diffusive behavior at long times. With the use of $N_\mathrm{c}=40$ collocation points, an error of about $15\%$ is incurred in the value of $\left\langle R^2\right\rangle$ and the slope of $g_1(t)$ vs $t$ in the subdiffusive regime, with respect to the results obtained with $N_\mathrm{c}=200$ collocation points. The need for a larger number of collocation points to attain improved convergence of the integrals in Eq. (\ref{eq:EVreg}) may be attributed to the increase in the magnitude of the Lennard--Jones repulsive energy relative to the energy of chain connectivity upon decreasing $k$. Moreover, Fig. \ref{fig:k3200} reveals an increase in chain extension on the scale of $\Delta s < N\delta$ upon decreasing $k$ from $75$ to $3$, indicating an increase in local chain stretching owing to the decreased penalty for stretching. The observed Rouse behavior may result from the concomitant increase in the spatial scale over which the chain exhibits flexible behavior. These features suggest that semiflexible behavior may be obtained for large $N$ by increasing $k$ or, alternatively, by decreasing the cutoff $\delta$. Entanglement effects, which are expected to yield subdiffusive behavior with a scaling exponent of $0.25$ for flexible chains,\cite{degennes,doiedwards} are not manifested in our results, possibly because the system sizes considered thus far are not sufficiently large.

\section{\label{sec:phantom} Phantom Chains}

In this section, we consider continuous chains that enforce topological constraints through interchain Lennard--Jones repulsion, but with the segments on a single chain interacting solely via elastic restoring forces. The absence of intrachain excluded volume interactions implies that only the first term on the right hand side of Eq. (\ref{eq:EVreg}) is retained. The chain properties described below have been calculated following the procedures described in Sec. \ref{sec:saw}.

\begin{figure}
\centering
\resizebox{80mm}{!}{\includegraphics{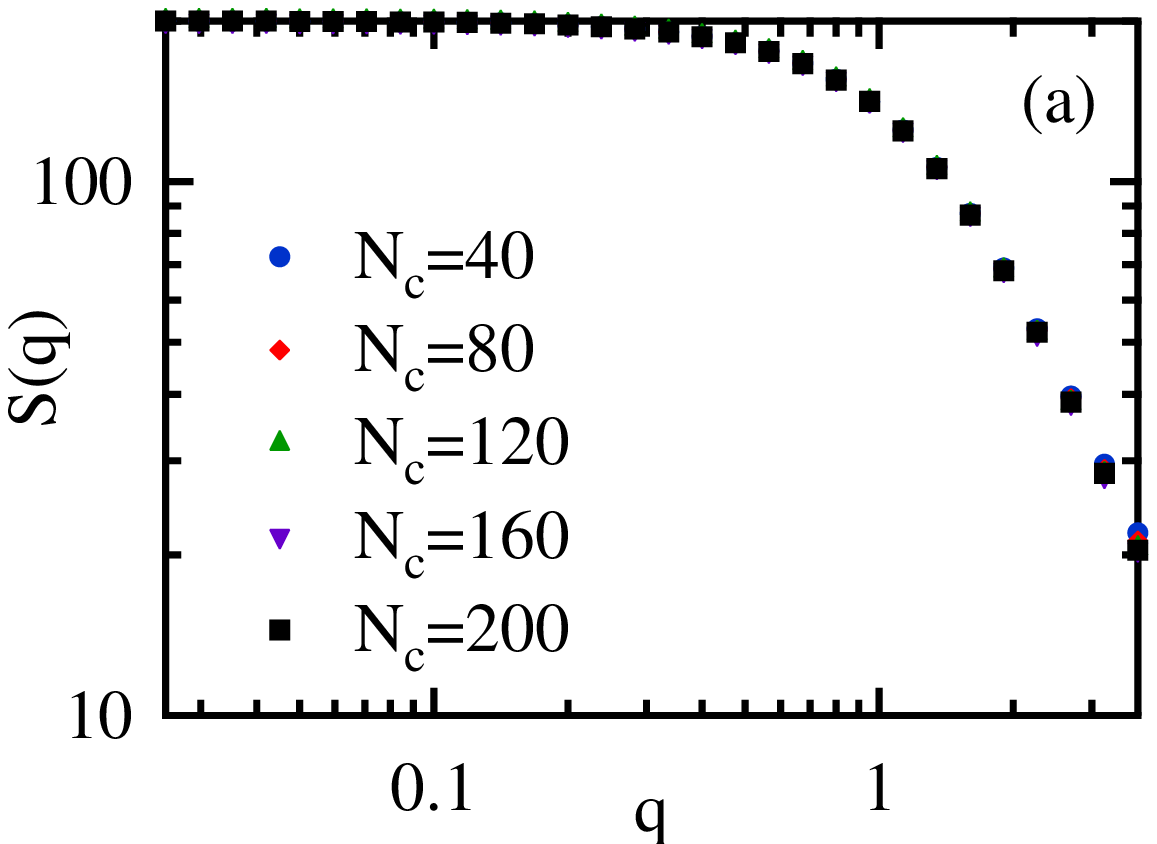}}
\vskip 1.25em
\centering
\resizebox{80mm}{!}{\includegraphics{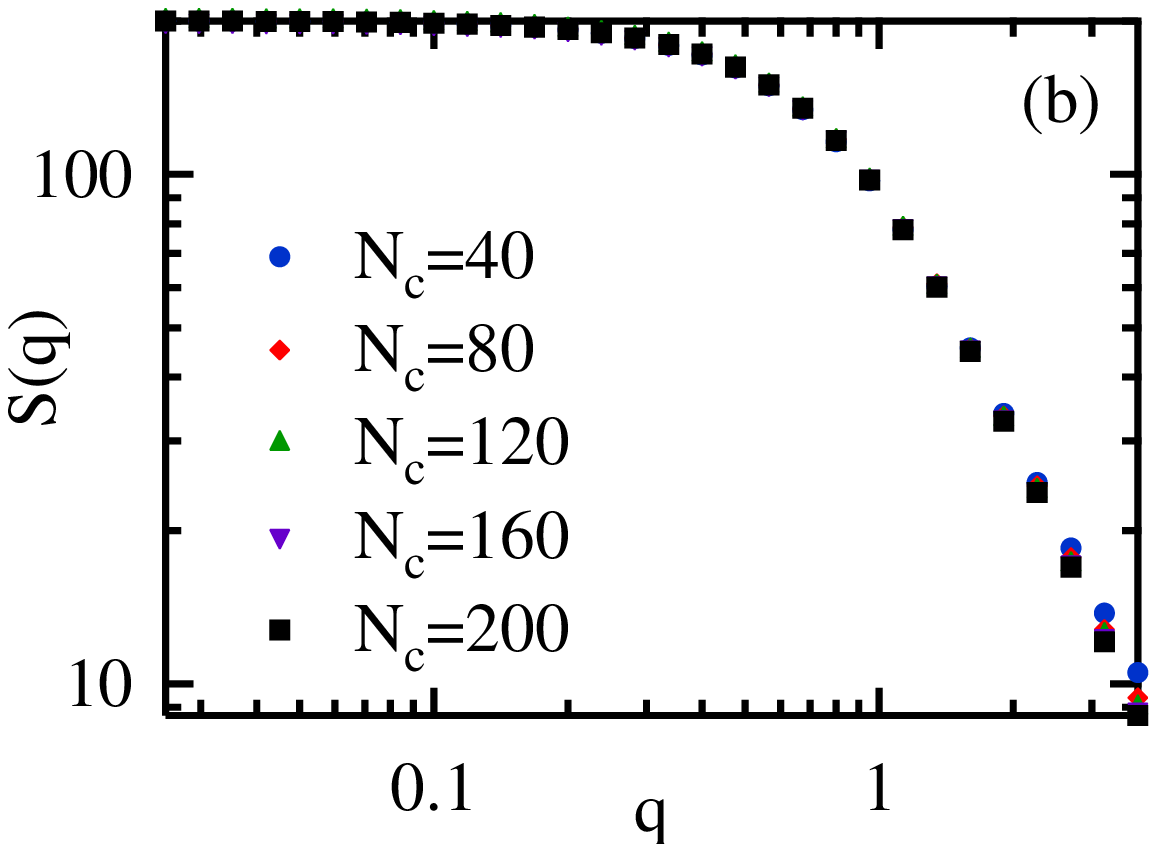}}
\caption[(Color online) Structure factor $S(q)$ for phantom chains of length $N=200$ using $N_\mathrm{c}=40,\ 80,\ 120,\ 160$ and $200$ collocation points with (a) $k=75$ and (b) $k=30$.]{\label{fig:sqnik2ni200}}
\end{figure}

\begin{figure}
\centering
\resizebox{80mm}{!}{\includegraphics{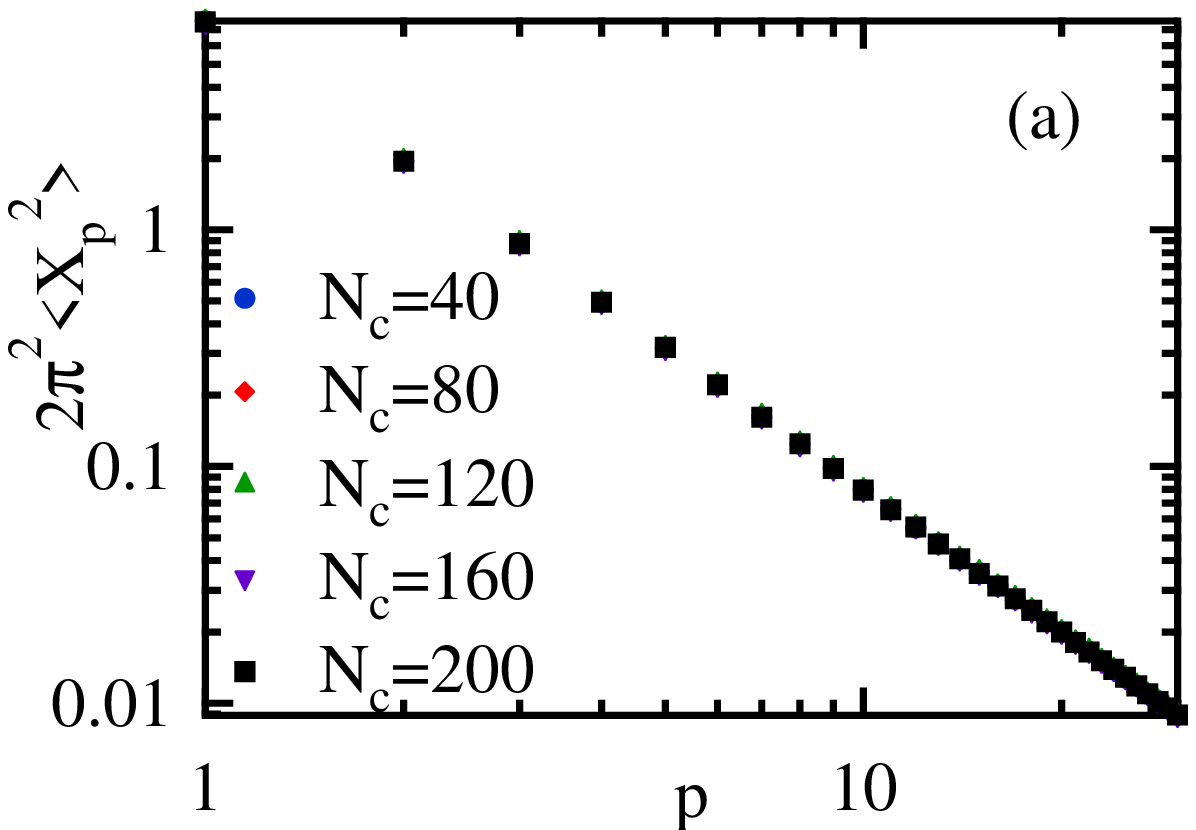}}
\vskip 1.25em
\centering
\centering
\resizebox{80mm}{!}{\includegraphics{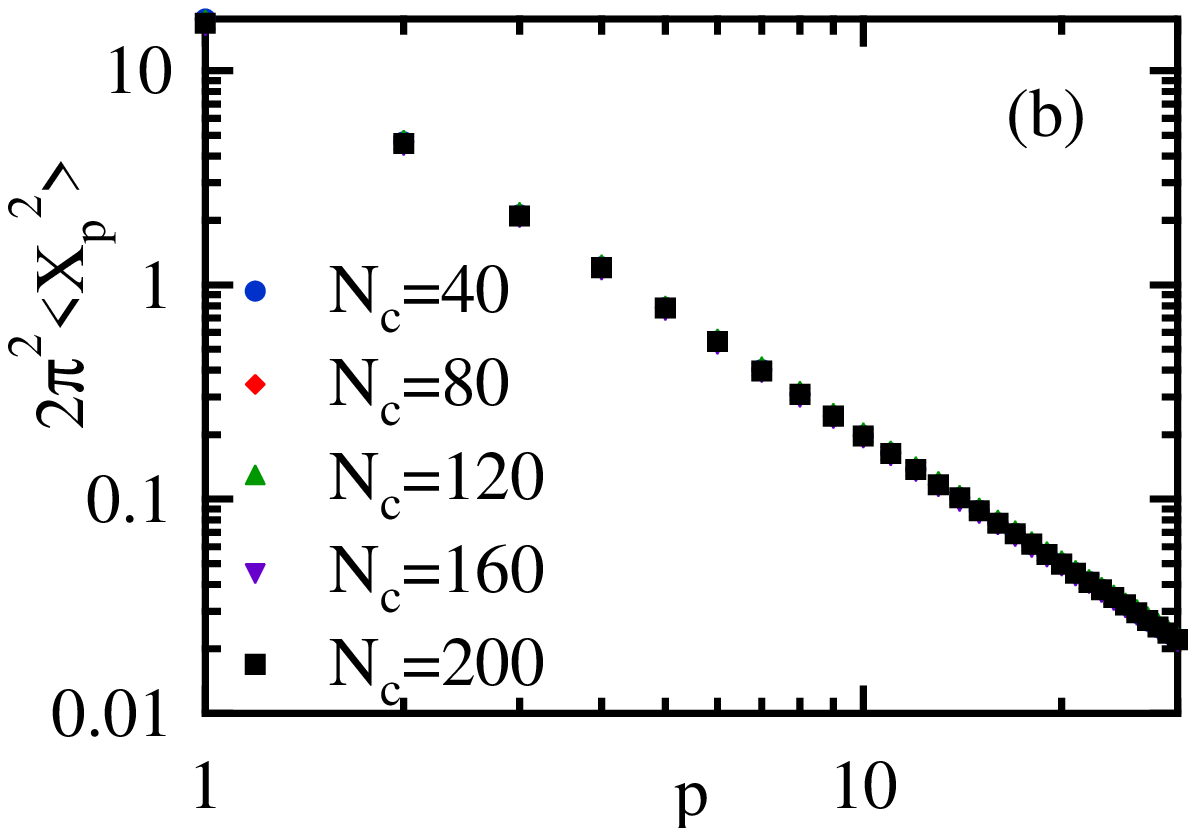}}
\caption[(Color online) Amplitude $<X_p(0)^2>$ of the $p^\mathrm{th}$ Rouse mode, normalized such that the data for $p=1$ coincide with the chain end-to-end distance for a Rouse chain, plotted as a function of $p$ for phantom chains of length $N=200$ using $N_\mathrm{c}=40,\ 80,\ 120,\ 160$ and $200$ collocation points with (a) $k=75$ and (b) $k=30$.]{\label{fig:xpsqnik2ni200}}
\end{figure}

Figure \ref{fig:sqnik2ni200} illustrates the structure factor for chains of length $N=200$ with $k=75$ and $k=30$. The slopes of $\log S(q)$ vs $\log q$ in the linear regime are found to be $-1.5 \ \pm 0.14$ and $-1.7 \ \pm 0.04$ corresponding to $k=75$ and $k=30$, respectively, approaching the predicted value of $-2$ for noninteracting flexible chains with increasing $k$. Figure \ref{fig:xpsqnik2ni200}, which depicts the mode amplitudes $\left \langle X_p(0)^2 \right\rangle$ vs $p$ for $N=200$ with $k=75$ and $k=30$, confirms the expected scaling $\left \langle X_p(0)^2 \right\rangle \sim 1/p^2$ for flexible chains.\cite{doiedwards} Clearly, the use of $N_\mathrm{c}=40$ collocation points is sufficient to yield convergence.

\begin{figure}
\resizebox{80mm}{!}{\includegraphics{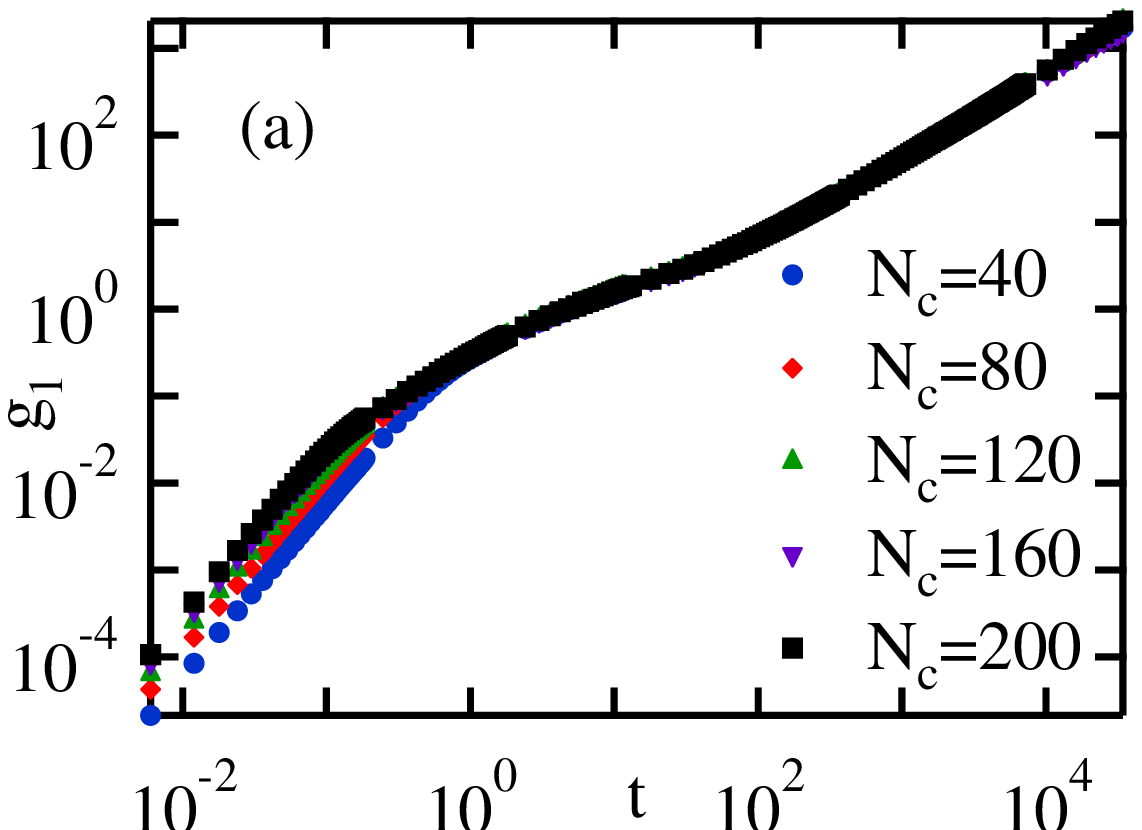}}
\vskip 1.25em
\centering
\resizebox{80mm}{!}{\includegraphics{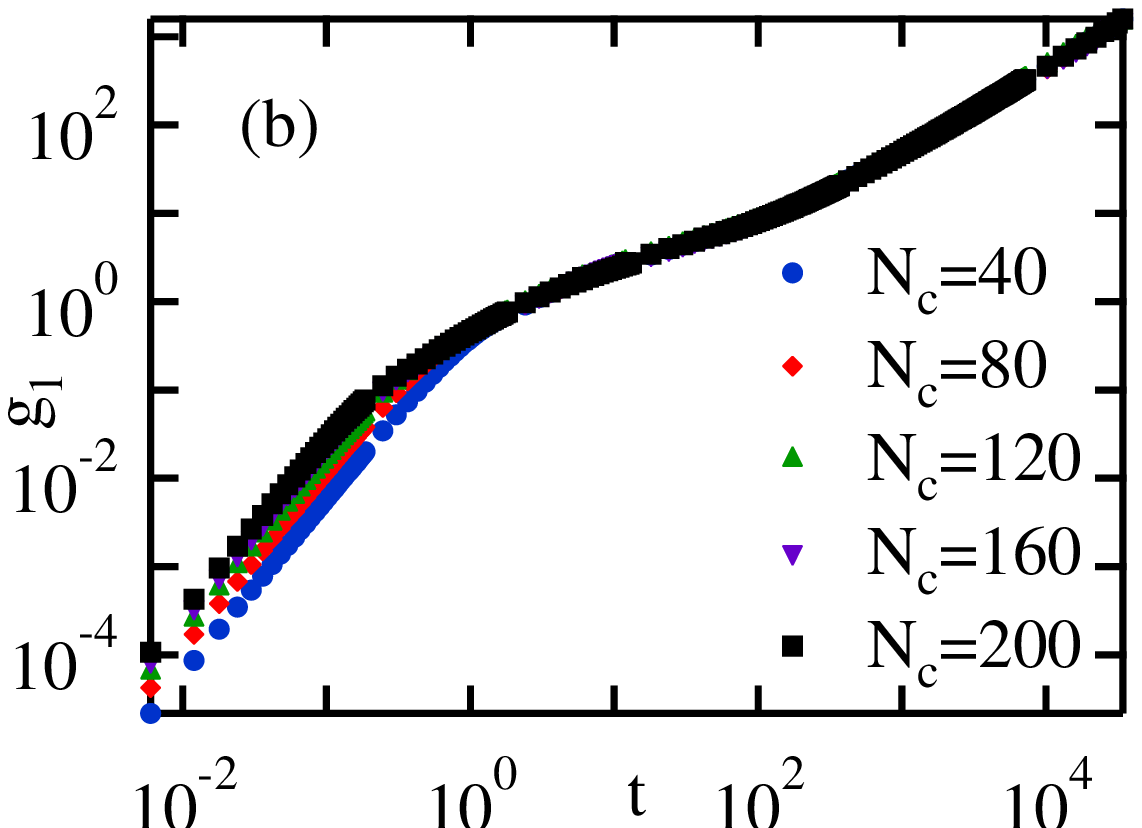}}
\caption[(Color online) Mean square displacement $g_1(t)$ at time $t$ averaged over the innermost $5\%$ of the chain for phantom chains of length $N=200$ using $N_\mathrm{c}=40,\ 80,\ 120,\ 160$ and $200$ collocation points with (a) $k=75$ and (b) $k=30$.]{\label{fig:msdnik2ni200}}
\end{figure}

\begin{figure}
\centering
\resizebox{80mm}{!}{\includegraphics{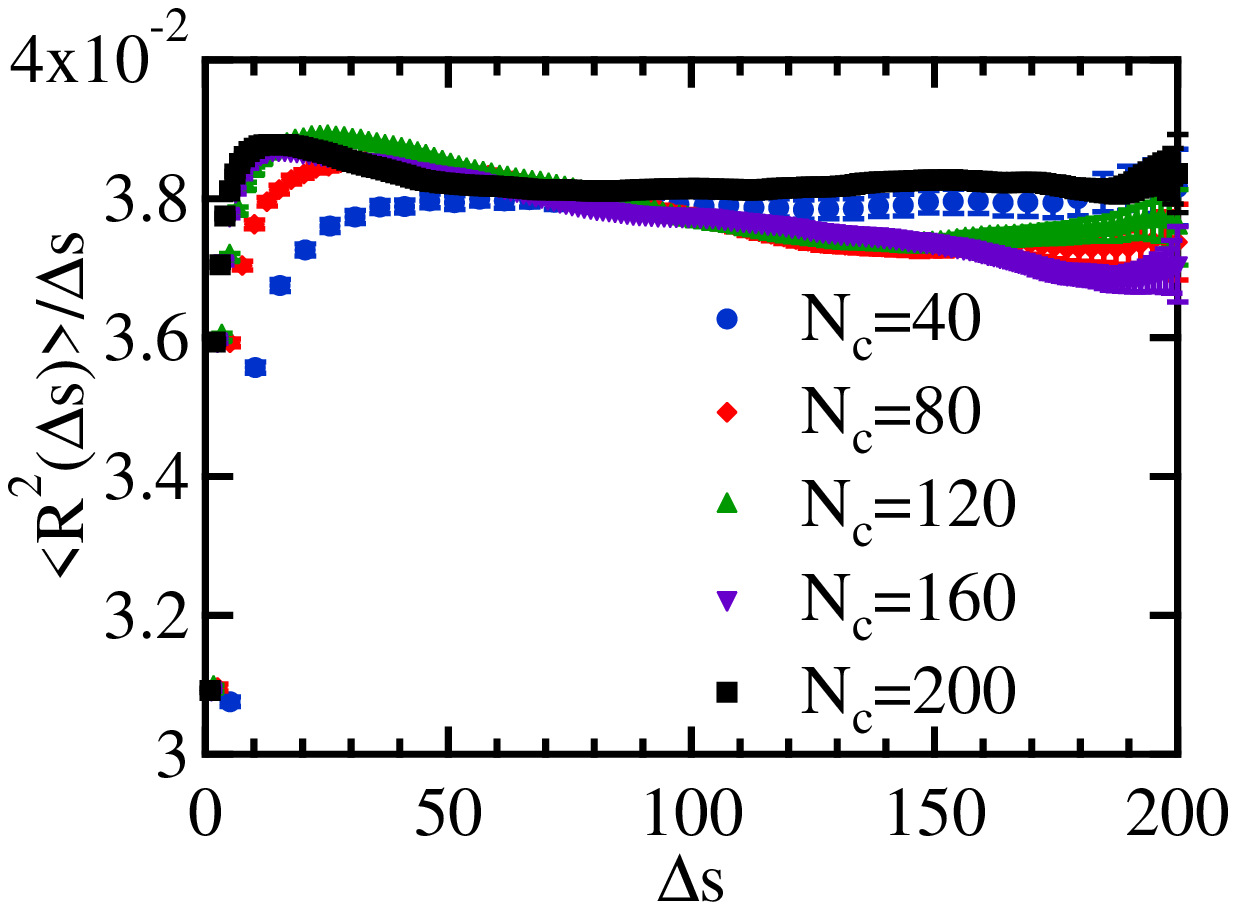}}
\caption[(Color online) Normalized mean square internal distance $\left \langle R^2(\Delta s) \right\rangle/\Delta s$ for phantom chains of length $N=200$ using $N_\mathrm{c}=40,\ 80,\ 120,\ 160$ and $200$ collocation points, with $k=75$. The error bars represent the standard error.] {\label{fig:msidni200}}
\end{figure}

The mean square displacement of the innermost $5\%$ of the chain is illustrated in Fig. \ref{fig:msdnik2ni200} for $N=200$ with $k=75$ and $30$. In both cases, a transition from an initial ballistic regime with $g_1(t) \sim t^2$ to a regime exhibiting the Rouse scaling $g_1(t) \sim t^{0.5}$ is observed, followed by normal diffusion at long times. Similar results (not shown here) are obtained for $N=100$ with $k=75$ and $30$, and $N=1000$ with $k=75$. The scaling $\tau_p \sim 1/p^2$ is also confirmed for the $N=1000$ system. In all cases, convergence is attained with $N_\mathrm{c}=40$ collocation points. We do not observe the scaling behavior $g_1(t) \sim t^{0.25}$ corresponding to the reptation regime for the system sizes studied, owing to the small chain sizes resulting from the use of relatively large values of $k$. For instance, as depicted in Fig. \ref{fig:msidni200}, chains of length $N=200$ with $k=75$ possess a mean square end-to-end distance $\left\langle R^2 \right\rangle$ of only about $8$.

\begin{figure}
\centering
\resizebox{80mm}{!}{\includegraphics{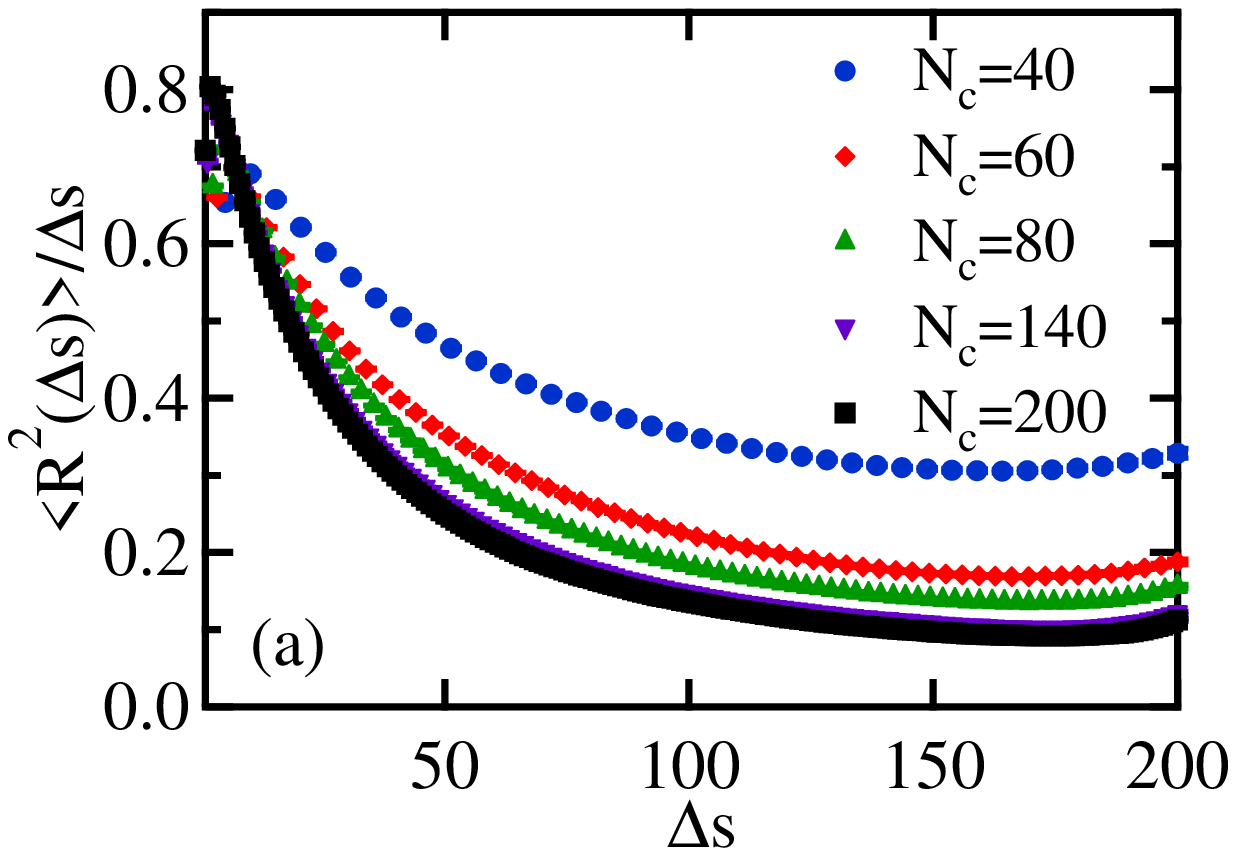}}
\vskip 1.25em
\centering
\resizebox{80mm}{!}{\includegraphics{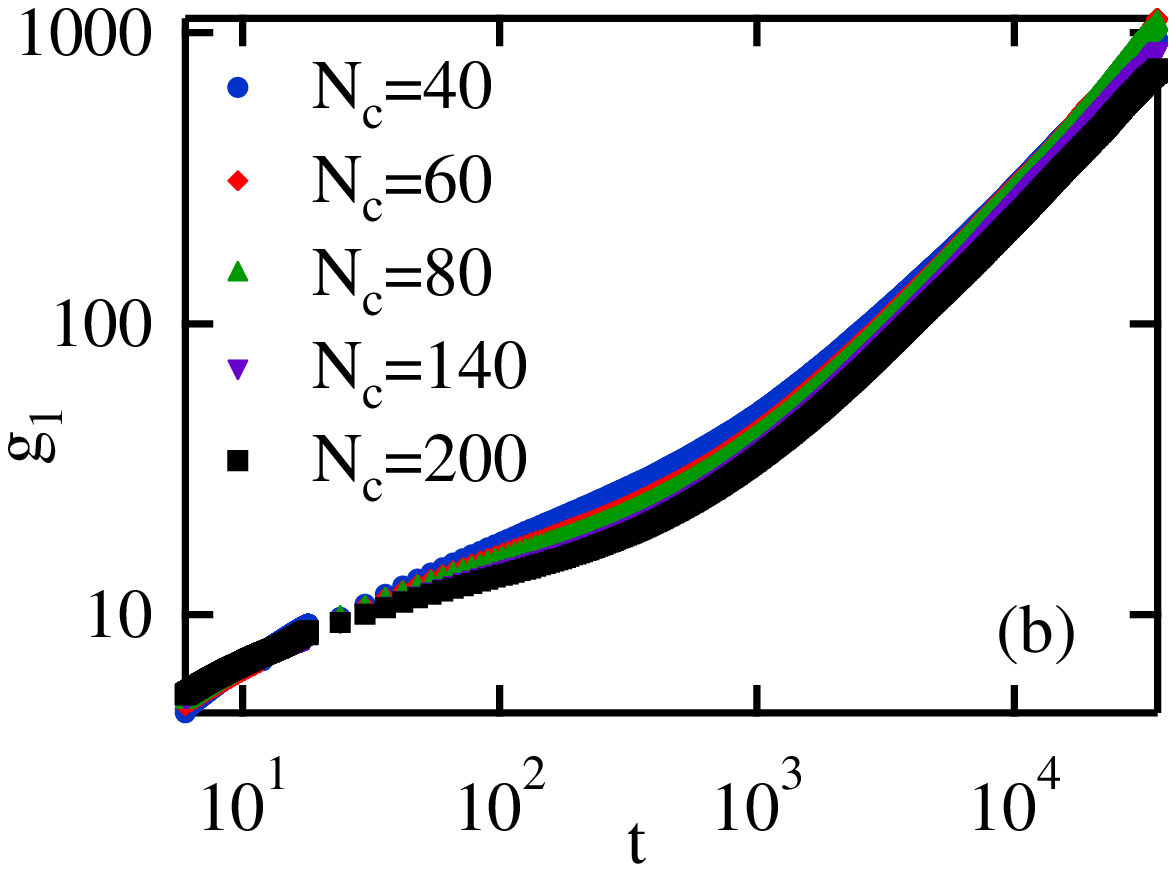}}
\caption[(Color online) Normalized mean square internal distance $\left \langle R^2(\Delta s) \right\rangle /\Delta s$ (a) and mean square displacement $g_1(t)$ averaged over the innermost $5\%$ of the chain (b) for phantom chains of length $N=200$ using $N_\mathrm{c}=40,\ 60,\ 80,\ 140$ and $200$ collocation points, with $k=3$. The error bars in (a) represent the standard error.] {\label{fig:k3ni200}}
\end{figure}

Results for a chain length of $N=200$ with $k=3$ are presented in Fig. \ref{fig:k3ni200}. Figure \ref{fig:k3ni200}(a) depicts the normalized mean square internal distance averaged over configurations separated by a time interval of $1\:080\tau$. The sampling interval was chosen to exceed the longest relaxation time $\tau_1$ determined from the single exponential decay of the corresponding mode correlation function by over a factor of $2$, to ensure independence of samples. The decrease in $\left\langle R^2 (\Delta s) \right \rangle / \Delta s$ as $\Delta s$ approaches $N$  may be the result of strong interchain excluded volume repulsion, since a decrease in $k$ signifies an increase in the strength of the Lennard--Jones repulsive energy in relation to the energy of chain connectivity. Consequently, a larger number of collocation points (at least $N_\mathrm{c}=80$) is necessary to achieve satisfactory convergence. A local increase in $\left\langle R^2 (\Delta s) \right \rangle / \Delta s$ for small $\Delta s$ concurrently appears as $k$ is decreased, on account of the reduced penalty for stretching. The ballistic regime resulting from correlated inertial motions has been omitted from Fig. \ref{fig:k3ni200}(b), which depicts the mean square displacement $g_1(t)$ vs $t$. Four regimes are observed over the time scales plotted in Fig. \ref{fig:k3ni200}(b). An initial regime in which $g_1(t) \sim t^{0.5}$ transitions into a regime spanning over a decade in time with the scaling behavior $g_1(t) \sim t^{0.25}$. This is followed by the scaling behavior $g_1(t) \sim t^{0.5}$. Finally, at long times, normal diffusion with $g_1(t) \sim t$ is observed. The observed scaling exponents coincide with the predictions of the reptation model for flexible chains.\cite{degennes,doiedwards} However, it is unclear whether the reptation picture is consistent with the observed shrinking of the chain at large scales, and anomalous subdiffusion may arise more generally from entropic barriers to chain motion.\cite{muthukumar} Overall, we see that the phantom chain model, derived by omitting the intrachain Lennard--Jones interactions, exhibits static and dynamic properties more consistent with entangled flexible chains, whereas the full model shows semiflexible characteristics.

\section{\label{sec:compt} Computational Time}

\begin{figure}
\centering
\resizebox{80mm}{!}{\includegraphics{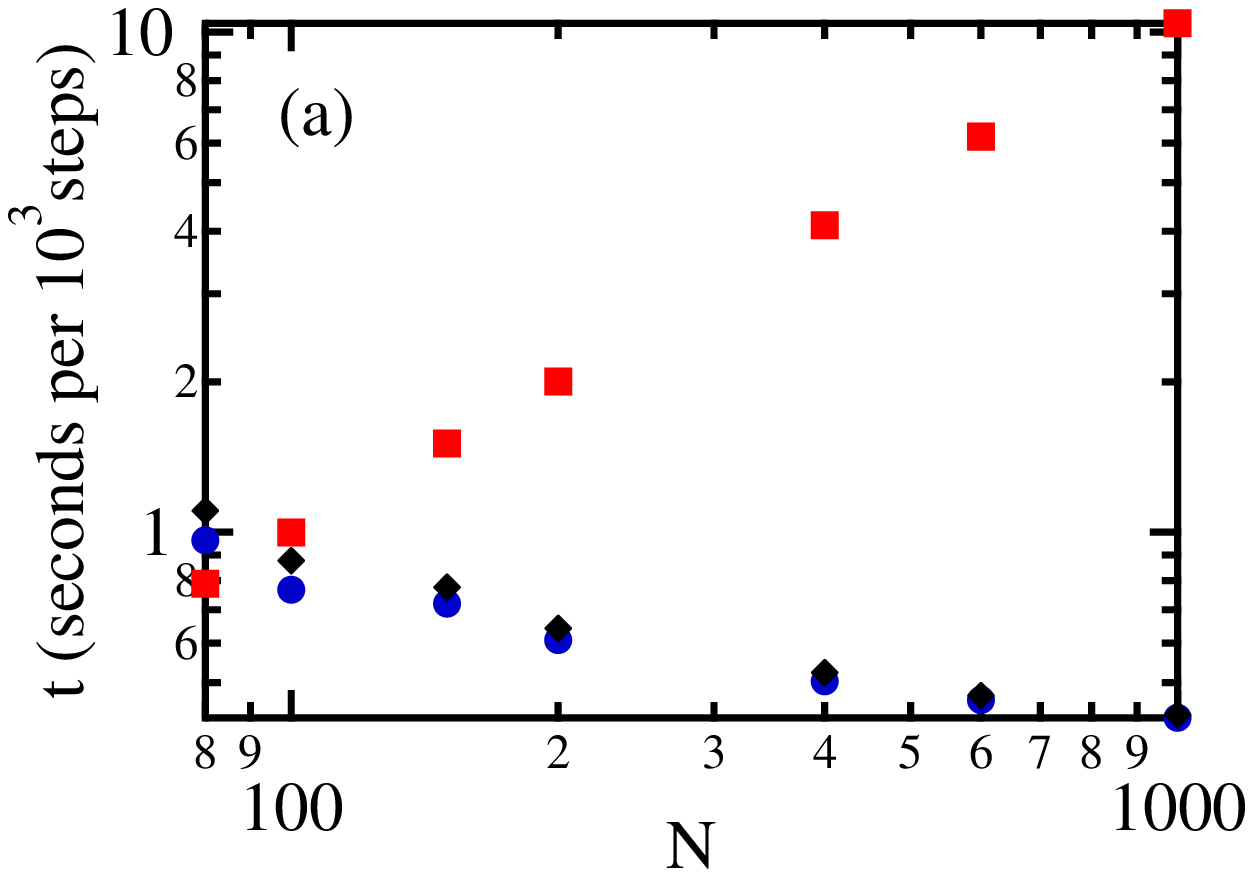}}
\vskip 1.25em
\centering
\resizebox{80mm}{!}{\includegraphics{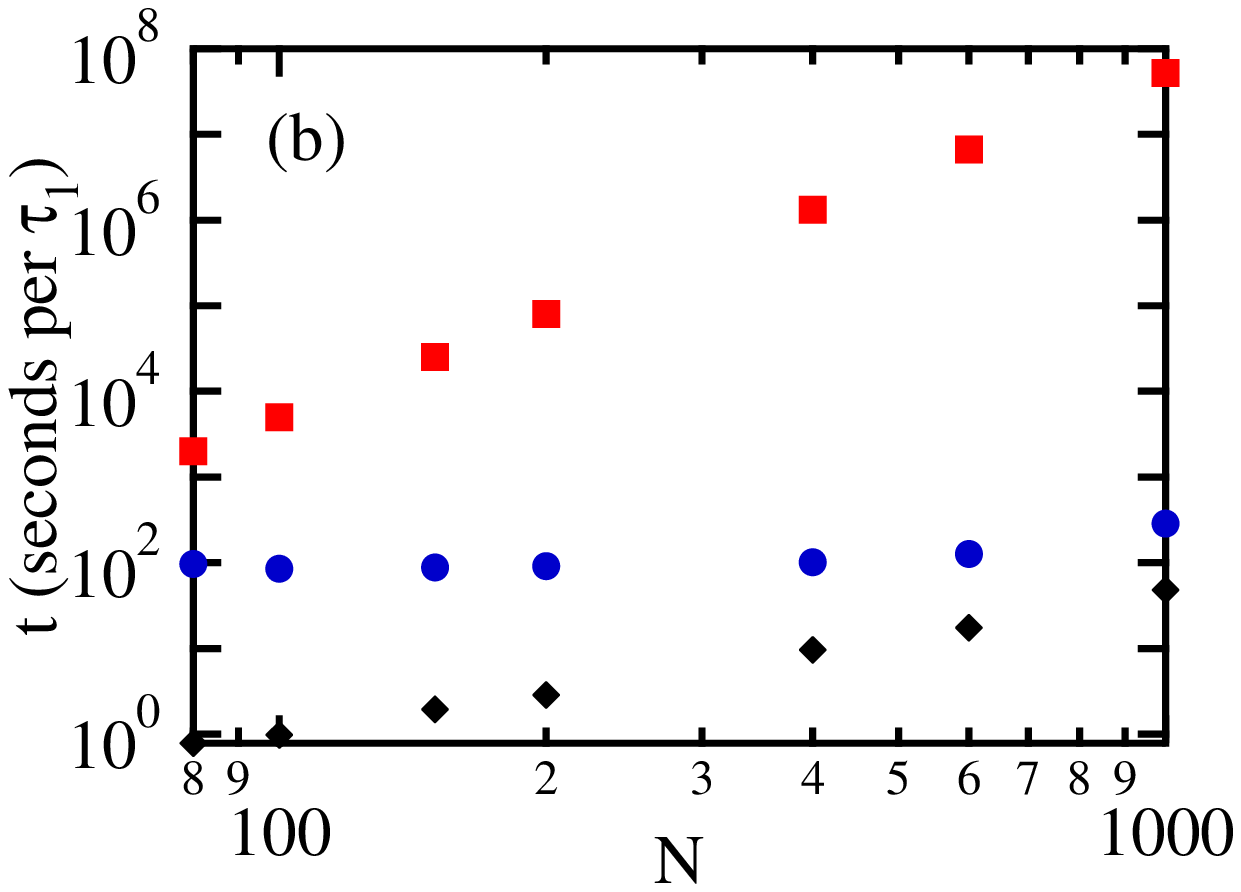}}
\caption[(Color online) CPU time required to simulate a system of $20$ FENE bead--spring chains \cite{kremergrest} (squares) and continuous chains in the presence (circles) and absence (diamonds) of intrachain repulsive interactions with $\rho=0.85$, $\delta=0.05$, $k=75$ and $N_\mathrm{c}=40$ per $10^3$ time steps (a) and per longest relaxation time $\tau_1$ (b) on a 2.83 GHz Intel Xeon processor.]{\label{fig:cput}}
\end{figure}

As demonstrated in the preceding sections, convergence of chain properties may be attained with the use of an $N$-independent number of collocation points, $N_\mathrm{c}$, which is governed by the values of $k$ and $\delta$. Figure \ref{fig:cput} provides a comparison of the CPU time required to simulate a system of $20$ chains of length $N$ at a monomer number density $\rho=0.85$ on a single $2.83$ GHz Intel Xeon processor based on the bead--spring model and the continuous chain model. The bead--spring model adopted is identical to that of Refs. \onlinecite{kremergrest} and \onlinecite{kremerprl}, and employs FENE springs with purely repulsive Lennard--Jones interactions among beads. The resulting root mean square distance of $0.97$ between adjacent beads on a chain  is comparable to the corresponding value of $0.94$ obtained upon replacing the FENE springs with stiff linear springs at a spring constant of $75$. The parameters $\delta=0.05$ and  $k=75$ are employed in the continuous chain model, and $N_\mathrm{c}=40$ collocation points are used. It is evident from Fig. \ref{fig:cput}(a) that the CPU time for a fixed number of steps required by the bead--spring model increases linearly with $N$, whereas the time taken by the continuous chain model remains almost unchanged with increasing $N$. The slight decrease in the CPU time required to simulate continuous chains as $N$ increases may be attributable to the fact that, at a constant segment number density $\rho = MN/V$ and at constant $M$, the volume of the periodic cell $V$ increases as $N$ is increased. Consequently, a larger volume is available to the system of chains discretized using a fixed number of collocation points $N_\mathrm{c}=40$, resulting in a decrease in the time required for building neighbor lists in simulations. We note, however, that with $M \sim N^{1/2}$ as in the preceding sections, $R_\mathrm{g}^3/V$ remains approximately constant as $N$ is increased at constant $\rho$, where $R_\mathrm{g} \sim N^{1/2}$ denotes the radius of gyration of an ideal chain. As a result, Lennard--Jones interactions among chains are adequately captured even for large $N$ with the use of an $N$-independent number of collocation points.

Figure \ref{fig:cput}(b) illustrates the CPU time required to simulate the system of chains for a time period corresponding to the longest relaxation time $\tau_1$ based on the bead--spring and continuous chain models. The value of $\tau_1$ for the bead--spring system with $N=100$ reported in Ref. \onlinecite{kremerprl}, in conjunction with the scaling relation $\tau_1(N) \sim N^3$ in the reptation regime, has been employed to determine the relaxation times for the bead--spring chains. The relaxation times for the continuous chain model were determined from the simulations detailed in Secs. \ref{sec:saw} and \ref{sec:phantom}. Two additional system sizes, namely, $N=400$ with $M=44$ and $N=600$ with $M=56$ were simulated to determine the corresponding values of $\tau_1$. In the absence of intrachain repulsive interactions, the relaxation times of chain lengths $N \leq 200$ could not be determined accurately from simulation, owing to the rapid decay of the mode correlation functions, and have been estimated by employing the scaling relation $\tau_1 \sim N^2$ for Rouse chains. It is apparent that the computational time required for simulations based on the continuous chain model grows relatively slowly in comparison to that required by the bead--spring model.

\section{\label{sec:conclude} Conclusions}

This study presents a new method for the molecular dynamics simulation of entangled polymer systems, wherein the polymer chains are represented as continuous Gaussian chains. The chains may interact via interchain and intrachain excluded volume interactions. Self-interactions are prohibited by imposing a cutoff $\delta$, whereby repulsive interactions between points on a single chain within a fractional distance of $\Delta\tilde{s} = \delta$ along the chain contour are forbidden. This approach is equivalent to assuming a locally flat intrachain repulsive potential within the short distance cutoff. Alternative approaches to regularizing the intrachain repulsive potential, not explored in this study, may also be adopted. For instance, the potential may be linearized at short range, thereby imposing an upper cutoff on the repulsive force, which may serve as an additional adjustable parameter to reproduce desired chain characteristics. It is demonstrated that the presence of intrachain repulsive interactions imparts correlations to the chain backbone, yielding behavior similar to that of semiflexible polymers, while flexible chain behavior is recovered in the absence of intrachain excluded volume interactions. The computational advantage of the method derives from the fact that the chain length $N$ appears only as a parameter in the equations governing the dynamics of the continuous chains [cf. Eq. (\ref{eq:MDcos})], thereby enabling the Fourier space representation of the dynamical equations to be solved pseudospectrally with a relatively small, $N$-independent number of collocation points. We establish that an $O(N)$ reduction in computational time per time step is attained in comparison with molecular dynamics simulations of the bead--spring model, and significant computational savings are achieved for chain sizes $N \gtrsim 100$.

In the present study, simulations based on the velocity Verlet method performed with a smaller time step of $\Delta t = 0.002\tau$ for the $N=1\:000$ system and the $N=200,\ k=3$ system were found to yield results identical to those obtained using a time step of $\Delta t = 0.006\tau$. While molecular dynamics simulations employing the Langevin thermostat permit the use of a larger time step of $\Delta t = 0.012\tau$,\cite{auhl} we have not explored the use of larger time steps in our simulations. We note that large values of $N$ may render the governing equations stiff [cf. Eq. (\ref{eq:MDcos})], entailing the use of small time steps for convergence. Semi-implicit schemes may alleviate the problem of stiffness and allow for relatively large time steps. We found that the implicit treatment of the second term on the right hand side of Eq. (\ref{eq:MDcos}) (with the evaluation of $\hat{\mathbf{R}}_i^q$ at time $t+\Delta t$) required a smaller time step of order $10^{-4}\tau$ for satisfactory accuracy in the absence of excluded volume interactions, although a time step of $\Delta t = 0.006\tau$ was found to suffice in the presence of excluded volume interactions. Therefore, although the dependence of the time step on $N$ is removed, such a semi-implicit scheme is not suitable for moderately large values of $N$. An alternative approach is to use a splitting scheme \cite{milstein} in conjunction with the analytical treatment of the linear terms in Eq. (\ref{eq:MDcos}) for large $N$, where the extra computational effort is offset by the savings resulting from the use of larger time steps.

The method presented in this study provides a framework for the incorporation of more general interaction potentials or applied forces, and extensions to membranes. For example, the bending potential of a semiflexible chain or a membrane may be taken into account explicitly. The model may be readily extended to provide a computationally tractable approach to the study of entangled polymer solutions with account for hydrodynamic interactions,\cite{dunweg} by an appropriate replacement of the scalar friction coefficient employed here with an inverse mobility tensor. The proposed approach is suitable for the investigation of a broad range of systems, including the structure and mechanical response of cellular networks \cite{mtosh1, mtosh2, mtosh3} and the rheology of entangled polymer solutions and melts.\cite{doiedwards, everaers}

\section*{Acknowledgments} This work was supported by the NSF grant DMR-CMMT-$\lambda$ 0904499 and made use of MRL Central Facilities supported by the MRSEC Program of the NSF under grant no. DMR-05-20415. The simulation code developed for this work was based in part on the LAMMPS molecular dynamics simulation code (http://lammps.sandia.gov) \cite{plimpton} and the FFTW package.\cite{frigo}

\clearpage
\pagebreak

\clearpage
\pagebreak

\clearpage
\pagebreak

\clearpage
\pagebreak

\clearpage
\pagebreak

\clearpage
\pagebreak

\clearpage
\pagebreak

\clearpage
\pagebreak

\clearpage
\pagebreak

\clearpage
\pagebreak

\clearpage
\pagebreak

\clearpage
\pagebreak

\clearpage
\pagebreak

\clearpage
\pagebreak
\listoffigures

\end{document}